\documentclass[%
superscriptaddress,
reprint,
nofootinbib,
amsmath,amssymb,amsbsy,aps,
prb,
longbibliography]{revtex4-1}

\usepackage[normalem]{ulem}
\usepackage{bbm}
\usepackage{verbatim}

\usepackage{cleveref}
\usepackage{braket}
\usepackage[pdftex]{graphicx}
\usepackage{epstopdf}
\usepackage{rotating}
\usepackage{dcolumn}
\usepackage{bm}
\usepackage{mathrsfs}
\usepackage{color}
\usepackage{leftidx}
\usepackage{bbm}
\usepackage{verbatim}
\usepackage{soul,xcolor}
\usepackage[toc,page]{appendix}
\usepackage{comment}
\usepackage{multirow} 
\usepackage[caption=false]{subfig}
\usepackage{extarrows}
\definecolor{greens}{rgb}{0,0.7,0}
\usepackage{amsmath}
\newcommand{\sfS}{{\sf S}}
\usepackage{lipsum}
\newcommand{\beginsupplement}{%
        \setcounter{table}{0}
        \setcounter{section}{0}
        \renewcommand{\thetable}{S\arabic{table}}%
        \setcounter{figure}{0}
        \renewcommand{\thefigure}{S\arabic{figure}}%
     }

\begin{document}

\title{\textcolor{black}{Theory of magnetostriction for multipolar quantum spin ice in pyrochlore materials}}
\author{Adarsh S. Patri}
\affiliation{Department of Physics and Centre for Quantum Materials, University of Toronto, Toronto, Ontario M5S 1A7, Canada}

\author{Masashi Hosoi}
\affiliation{Department of Physics, University of Tokyo, 7-3-1 Hongo, Bunkyo, Tokyo 113-0033, Japan}

\author{SungBin Lee}
\affiliation{Department of Physics, Korea Advanced Institute of Science and Technology, Daejeon, 34141, Korea}

\author{Yong Baek Kim}
\affiliation{Department of Physics and Centre for Quantum Materials, University of Toronto, Toronto, Ontario M5S 1A7, Canada}
\affiliation{School of Physics, Korea Institute for Advanced Study, Seoul 02455, Korea}

\setstcolor{red}

%\date{\today}

\begin{abstract}
Multipolar magnetism is an emerging field of quantum materials research. The building blocks of multipolar phenomena are magnetic ions with a non-Kramers doublet, where the orbital and spin degrees of freedom are inextricably intertwined, leading to unusual spin-orbital entangled states. The detection of such subtle forms of matter has, however, been difficult due to a limited number of appropriate experimental tools. In this work, motivated by a recent magnetostriction experiment on Pr$_2$Zr$_2$O$_7$, we theoretically investigate how multipolar quantum spin ice, an elusive three dimensional quantum spin liquid, and other multipolar ordered phases in the pyrochlore materials can be detected using magnetostriction. We provide theoretical results based on classical and/or quantum studies of non-Kramers and Kramers magnetic ions, and contrast the behaviors of distinct phases in both systems. Our work paves an important avenue for future identification of exotic ground states in multipolar systems.

\end{abstract}

\maketitle

\section{Introduction}

The development of a robust understanding of emergent phenomena in strongly correlated quantum systems\cite{review_www_ybk_2014,review_schaffer_ybk_2016} ultimately requires deep insight into the many-body ground state, and the excitations it can support.
This historically successful paradigm has been challenged in recent times by \textcolor{black}{two prominent} examples: quantum spin liquids (QSLs) and multipolar ordered-states (MPOs).
QSLs, which are long-range entangled correlated paramagnets, support deconfined fractionalized excitations (spinons) that \textcolor{black}{couple to an emergent gauge field}\cite{Gingras_2014_qsi_review,balents_sl}.
QSLs arise from a variety of mechanisms, from geometrical frustration\cite{rau_gingras_frustration,gingras_bramwell_science_review} to anisotropic bond-dependent interactions\cite{kitaev_model, kitaev_rau_kee_review, sl_review}.
The lack of magnetic ordering presents an obvious challenge as to its detection with conventional probes, and despite efforts from neutron scattering \cite{kimura_nakatsuji_ncomm_pzo_2013, broholm_pzo_2017}, a true smoking-gun signature has proven to be elusive.
Analogously, MPOs also defy detection by conventional probes, despite falling under the purview of spontaneous symmetry breaking.
These ordered states, which arise from spin-orbit coupling and crystalline electric fields (CEFs) placing restrictions on localized electron orbitals' shapes, do not possess just a simple dipolar moment.
Instead, they support non-trivial charge and magnetic density distributions (described by higher-rank multipolar moments\cite{multupole_rev_1,multupole_rev_3}), which fail to directly couple to neutrons and other probes of ordering, and have been appropriately named ``hidden orders''.

The central question that remains for both of these phenomena is: how can the existence and properties of QSLs' and MPOs' interacting many-body ground states be examined if they shy away from conventional probing tools?
The answer to this question has had some success in recent times where novel elastic-based techniques in multipolar heavy fermion systems seem to indicate the onset of MPOs \cite{ultrasound_v, elasto_multipolar, magnetostiction_expt_ir, wartenbe_magnetoelastic_uru2si2_2019}.
Motivated by such experiments, we ask: could the experimentally reticent QSLs be exposed if they arise from interacting multipolar moments?

The pyrochlore oxide family provides a unique setting for the closer examination of the posed question. 
In these compounds, local CEFs result in low-lying Kramers or non-Kramers ground states\cite{ross_savary_prx}.
The Kramers ions typically host dipolar moments, and as such are more receptive to conventional probing tools\cite{savary_order_by_disorder, Yasui_ybto_2003, Chang_ybto_2012, Armitage_ybto_2015, gaudet_ybto_2016, broholm_ybto_111_2017, Broholm_ybto_mm_2019}.
However, the non-Kramers doublet found in Pr$_2$Zr$_2$O$_7$ hosts time-reversal even electric quadrupolar moments (and an accompanying magnetic dipolar moment)\cite{onada_pesuospin_model_1,onada_pesuospin_model_3}.
These moments reside on a pyrochlore lattice, \textcolor{black}{where frustrated pairwise interactions allow the possible existence of a type of QSL with an emergent U(1) gauge field and accompanying bosonic spinons, known as quantum spin ice\cite{hermele_pyrochlore_photons, isakov_ybk_hcb_2008, savary_gmft, sbl_gmft, kimura_nakatsuji_ncomm_pzo_2013, petit_afq_pzo_2016,savary_balents_disorder_qsl_2017,matsuda_qsl_2018}.}

In this work, we propose that magnetostriction (length change under an external magnetic field) provides a sharp and distinct signature of quantum spin ice formed from non-Kramers multipolar moments. 
The current work is directly motivated by a recent experiment on the quantum spin ice candidate material, Pr$_2$Zr$_2$O$_7$ \cite{nan_tang_aps_2019}.
\textcolor{black}{In order to validate this proposal, we contrast its difference by presenting the distinctive length change behaviours of possible ordered states in Kramers/non-Kramers ions.} 
In doing so, we establish a comprehensive theory of magnetostriction for a number of possible novel emergent phases in both non-Kramers and Kramers pyrochlore systems.
\textcolor{black}{The} findings are based on corroborating classical analysis and exact diagonalization of quantum models on the pyrochlore lattice.
\textcolor{black}{Our theoretical results provide a means to identify the existence of a QSL's many-body ground state, and to distinguish the various ordered phases in the pyrochlore family.} 
Our work lays the foundations upon which targeted experimental investigations can be conducted on QSLs, and provides a new direction of inquiry in the field of multipolar magnetism.

\section{Pseudospin-1/2 model of pyrochlore materials}
\label{sec_spin_model}
%%%
\begin{figure}[t]
\includegraphics[width=0.95\linewidth]{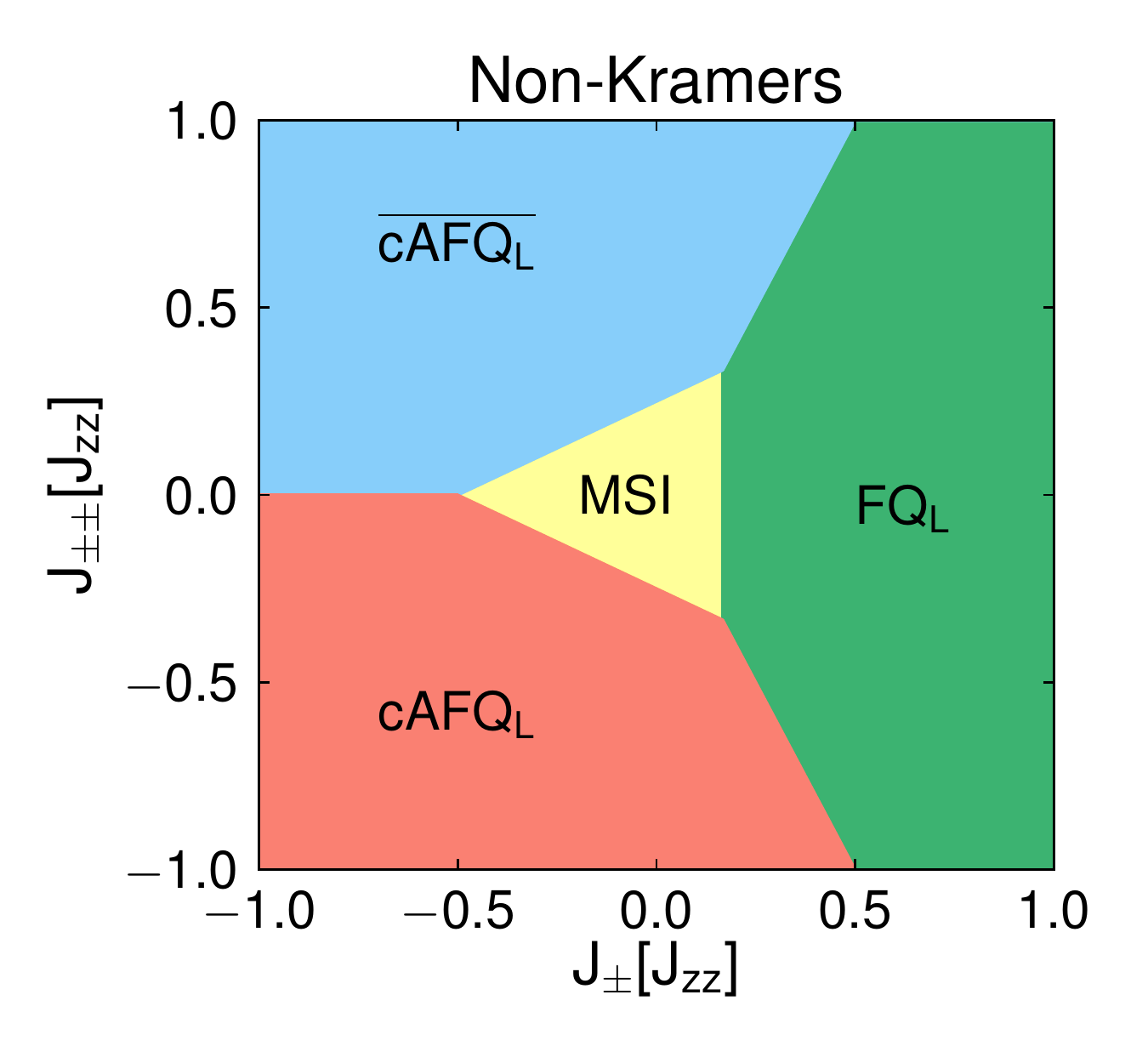}
\caption{ 
Classical Phase diagram of Eq. \ref{eq_hamiltonian_all} for non-Kramers ions ($J_{z\pm}=0$). The depicted phases are multipolar spin ice (MSI) of $T_{1g}$, coplanar anti-ferroquadrupolar (cAFQ$\rm{_L}$) of $T_{2g}$,  a second coplanar anti-ferroquadrupolar ($\rm{\overline{cAFQ_L}}$) of $T_{1g}$, and ferro-quadrupolar ($\rm{FQ_L}$) of $E_g$ symmetry. Here we use the subscript `$\rm{L}$' to indicate orderings in the local basis (Supplementary Information I and XIII).
cAFQ$\rm{_L}$ and $\rm{\overline{cAFQ_L}}$ are related to each other by a local $C_{4z}$ rotation on each sublattice.}
\label{fig_classical_phases}
\end{figure}
%%%
In the R$_2$M$_2$O$_7$ oxide family, the local moments arise from the $f$ electrons of the R$^{3+}$ rare-earth ions. 
Importantly, the surrounding cage of O$^{2-}$ ions subject the $f$ electrons to a local $D_{3d}$ crystalline electric field (CEF), which splits the $J$ multiplet of isolated R$^{3+}$ ions to yield low-lying Kramers or non-Kramers doublet ground states, depending on the nature of the R$^{3+}$ ions. 
\textcolor{black}{In Yb$_2$Ti$_2$O$_7$\cite{Hodges_2001, Yaouanc_ybto_2016,broholm_ybto_111_2017, Armitage_ybto_2014, thompson_ybto_2017}, for example, the $J=7/2$ multiplet is split to yield Kramers ground states that support conventional magnetic dipole moments, which can be efficiently represented by the pseudospin-1/2 operator $\bm{ \sfS} = \bm{J}$.}
In the intriguing candidate quantum spin-ice material, Pr$_2$Zr$_2$O$_7$, the $J=4$ degenerate manifold is partially lifted to yield non-Kramers (doublet) ground states of an even number of $f$ electrons. As a consequence, these support, in addition to a conventional magnetic dipole moment, more exotic time-reversal even quadrupolar moments. 
The multipolar moments can be efficiently represented by the pseudospin components, ${\sfS}^x = \overline{J^x J^z}$, $\sfS ^y = \overline{J^y J^z}$, and $\sfS ^z = J^z $, where the overline indicates a symmetrized product\cite{onada_pesuospin_model_1}.
In the local frame of each sublattice (as described in Supplementary Information I), members of the R$_2$M$_2$O$_7$ family obey the following generic nearest-neighbour pseudospin-1/2 model\cite{onada_pesuospin_model_1,onada_pesuospin_model_2,onada_pesuospin_model_3},
%%%
\begin{align}
 \label{eq_hamiltonian_all}
  {\mathcal{H}} = &\sum_{\langle i j \rangle} \Bigg \{ J_{zz} \sfS_i^{z} \sfS_j^{z} - J_{\pm} \left( \sfS_i^{+} \sfS_j^{-}  + \sfS_i^{-} \sfS_j^{+}    \right) \nonumber \\
 & + J_{\pm \pm}	\left(   \gamma_{ij} \sfS_i^{+} \sfS_j^{+} +  \gamma_{ij}^* \sfS_i^{-} \sfS_j^{-}    \right)  \\
 & + J_{z \pm}	\left[ \sfS_i^{z} \left(   \zeta_{ij}  \sfS_j^{+} +  \zeta_{ij}^* \sfS_j^{-} \right) + \left( \zeta_{ij} \sfS_i^{+}  + \zeta_{ij}^* \sfS_i^{-}  \right) \sfS_j^z \right] \Bigg\} \nonumber
\end{align}
%%%
where $\sum_{\langle i j \rangle} $ is a sum over nearest neighbour sites $i$ and $j$ on the pyrochlore lattice, $\gamma_{ij}$ and $ \zeta_{ij} = -  \gamma_{ij}^*$ are unimodular complex numbers listed in Supplementary Information I.
$J_{zz}>0$ is an antiferromagnetic interaction that gives rise to the celebrated two-in, two-out spin ice rule.
Due to the quadrupolar nature of $\sfS^{x,y}$ in the non-Kramers variety, $J_{z \pm}=0$ by time-reversal symmetry. 
\textcolor{black}{We note that, in anticipation of the discussion to follow, the exchange coupling constants are conventionally taken to be independent of applied magnetic fields, as the ground state doublet in Pr$_2$Zr$_2$O$_7$ is well separated from its excited states \cite{kimura_nakatsuji_ncomm_pzo_2013}.}

Figure \ref{fig_classical_phases} presents the $T=0$ classical phase diagram associated with Eq. \ref{eq_hamiltonian_all} with $J_{z \pm}=0$.
This classical non-Kramers phase diagram provides a zoo of possible phases: a classical 2-in, 2-out multipolar spin ice (MSI) phase of $T_{1g}$ symmetry, a coplanar anti-ferroquadrupolar (cAFQ$\rm{_L}$) phase with $T_{2g}$ symmetry, another coplanar antiferro-quadrupolar ($\rm{\overline{cAFQ_L}}$) of $T_{1g}$ symmetry, and a ferro-quadrupolar ($\rm{FQ_L}$) of $E_g$ symmetry\cite{shannon_classical}. 
Here we use the subscript `$\rm{L}$' to indicate orderings in the local basis.

Although we have so far discussed classical multipolar phases, it is highly suggestive from parton mean-field theory (known as gauge mean field theory, gMFT\cite{savary_gmft, sbl_gmft}) studies that quantum phases are the descendants of these parent classical phases. 
Indeed, in gMFT, the classical SI phase gets promoted to a U(1) quantum spin-liquid phase, which is characterized by the existence of deconfined bosonic spinons (magnetic monopoles, in the spin ice literature \cite{sondhi_csi_monopoles}) coupled to a U(1) gauge field\cite{hermele_pyrochlore_photons}. The other classically ordered phases get promoted to Higgs phases in gMFT, where the bosonic monopole condenses thus eliminating the emergent gauge field.
In the current work, instead of using gMFT, we examine the quantum model using exact diagonalization, which indeed confirms the relevant phase diagram (Supplementary Information XI).

\section{Elastic Strain Coupling to Local Moments}
In this section, we examine the coupling of the local R$^{3+}$ moments to the elastic normal modes.
The cubic nature of the underlying Bravais lattice constrains (by $O_h$ point group) the elastic energy to be of the form,
%%%%%%
\begin{equation}
\begin{aligned}
\mathcal{F}_{\text{lattice}} &= \frac{c_{B}}{2} \left(\epsilon_B^2 \right) + \frac{c_{11} - c_{12}}{2} \left( \epsilon_{\mu} ^2 + \epsilon_{\nu} ^2 \right)  \\
& + \frac{c_{44}}{2} \left( \epsilon_{xy}^2 + \epsilon_{yz} ^2 + \epsilon_{xz} ^2 \right)  ,
\label{F_lattice_normal}
\end{aligned}
\end{equation}
%%%%%%
where the crystal's deformation is described by the components of the strain tensor $\epsilon_{ik}$, and $c_{ij}$ is the elastic modulus tensor describing the stiffness of the crystal. Here $c_B$ is the bulk modulus,  $\epsilon_B \equiv \epsilon_{xx} + \epsilon_{yy} + \epsilon_{zz}$ is the volume expansion of the crystal, $\epsilon_{\nu} \equiv (2 \epsilon_{zz} - \epsilon_{xx} - \epsilon_{yy}) / \sqrt{3}$ and $\epsilon_{\mu} \equiv (\epsilon_{xx} - \epsilon_{yy})$ are cubic normal mode lattice strains.
Due to the sublattice nature of pyrochlore lattice, we specify that the elastic strain tensors, magnetic fields, as well as local moments, written in the local basis of a given sublattice-$\alpha$ possess a sublattice index i.e. $\sfS_{\alpha}^{x,y,z}$,  $h_{\alpha }^{x,y,z}$, $\epsilon_{ij}^{\alpha}$. In the global basis $\{ (1,0,0) , (0,1,0), (0,0,1) \}$, we write the same quantities without the sublattice index. 
In coupling the elastic strain to the localized moments, we enforce the local $D_{3d}$ point group symmetry of the surrounding CEF locally.
We present in Supplementary Information II, the relationships between local and global quantities and their transformations under $D_{3d}$.

Firstly, since the non-Kramers XY pseudospin components are time-reversal even electric quadrupolar moments, they can couple linearly to elastic strain as,
%%%
\begin{align}
\mathcal{F}_{\text{XY,NK}}  = & - k_1 \Big[ \sfS_{\alpha}^x \left(\epsilon_{xx}^{\alpha} - \epsilon_{yy}^{\alpha} \right)  - 2 \sfS_\alpha ^{y}  \epsilon_{xy} ^{\alpha}  \Big] \nonumber \\
&- {k_2  \Big[ \sfS_\alpha ^{x } \epsilon_{xz} ^{\alpha}   + \sfS_\alpha ^{y } \epsilon_{yz} ^{\alpha}   \Big]}, 
\label{eq_coupling_XY_NK}
\end{align}
%%%
where we have introduced Einstein summation notation for $\alpha$. We explicitly denote the non-Kramers case by the subscript NK, and $k_{1,2}$ are phenomenological coupling constants.
%The XY couplings for the Kramers case is discussed later in Supplementary Information IX.
The Z pseudospin component contains the time-reversal odd magnetic dipole moment and can only couple to the elastic strain in the presence of a time-reversal breaking external magnetic field, $\bm{h}$, to yield,

\begin{align}
\label{eq_coupling_Z}
\mathcal{F}_{\text{Z}}  = & - g_1 \sfS_\alpha ^{z} \Big[ \left(\epsilon_{xx}^{\alpha} - \epsilon_{yy}^{\alpha} \right) h_\alpha ^{x} - 2 \epsilon_{xy} ^{\alpha} h_\alpha ^{y} \Big]  - {g_4 \sfS_\alpha ^{z} h_\alpha ^{z}\Big[\epsilon_{zz}^{\alpha}   \Big] } \nonumber \\
& - g_2 \sfS_\alpha ^{z} \Big[ \epsilon_{xz} ^{\alpha} h_\alpha ^{x}   + \epsilon_{yz} ^{\alpha} h_\alpha ^{y}    \Big] - g_3 \sfS_\alpha ^{z} h_\alpha ^{z} \Big[ \epsilon_{xx}^{\alpha} + \epsilon_{yy}^{\alpha} \Big]  
\end{align}
%%%
where $g_{1,2,3,4}$ are coupling constants, and we again employ Einstein summation notation for $\alpha$.
Equation \ref{eq_coupling_Z} is common to both non-Kramers and Kramers ions.

%%%
\begin{figure*}[t]
\centering
\includegraphics[width=1\linewidth]{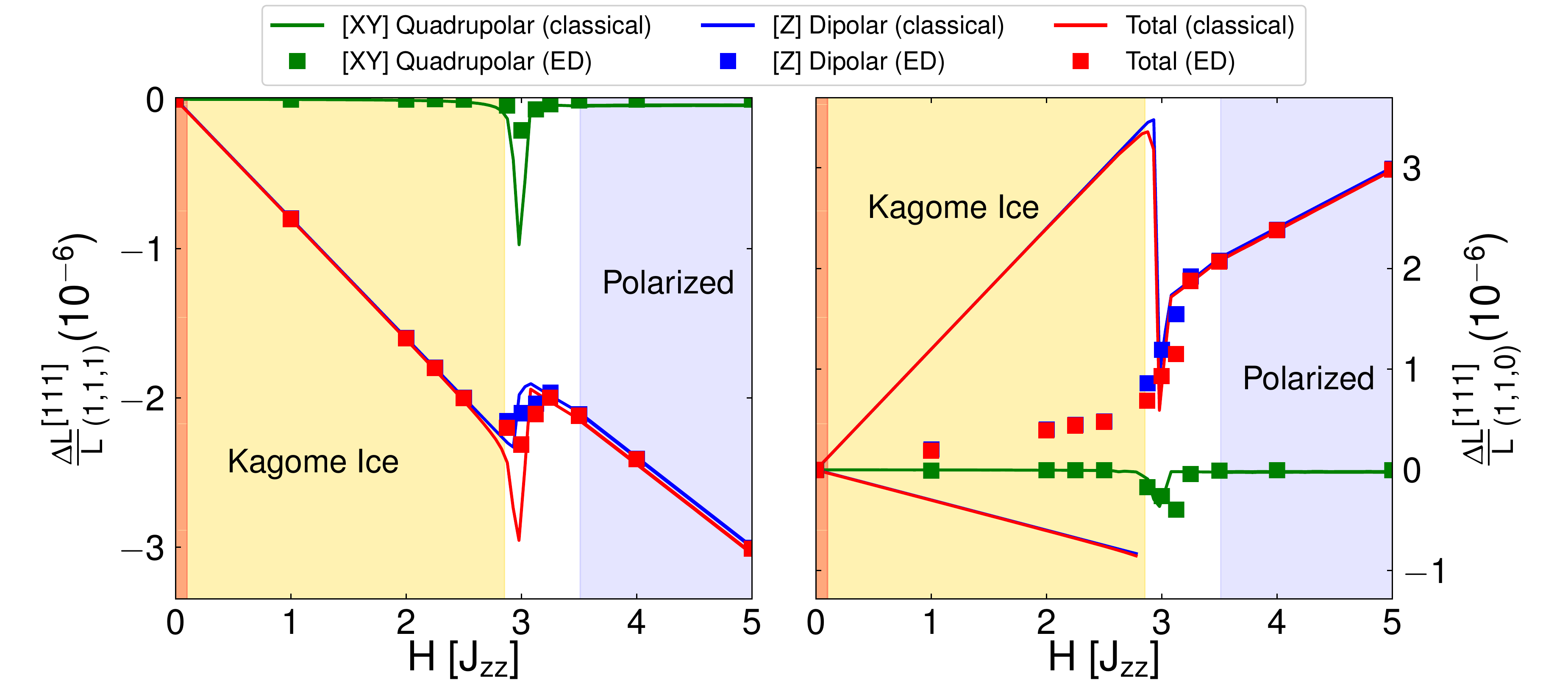}
\caption{ Length change, $\frac{\Delta L}{L}$, under applied [111] magnetic field, ${\bm{h}}$ for spin ice phase $J_{\pm} = 0.02 J_{zz}, \  J_{\pm \pm} = 0.05 J_{zz}$. Left: along the (1,1,1) direction, Right: along the (1,1,0) direction. For an infinitesimal field, the classical spin ice enters into the degenerate Kagome spin ice phase (denoted by yellow shaded region).
\textcolor{black}{The quantum spin ice is, however, stable for a small window of magnetic field strengths\cite{onada_qmc}; we schematically denote this region by the orange shaded region (the size of the region is amplified for ease of viewing).}
 For large fields, the system resides in a fully-polarized state (indigo-blue shaded region). The {green}, {blue} and {red} curves [squares] denote the length change arising from the XY pseudospin (quadrupolar), Z pseudospin (dipole), and combined contributions, respectively, from classical [32 site exact diagonalization] studies. 
\textcolor{black}{The dipole plot-line is (slightly) purposely shifted from the total length change plot-line to more easily visualize the individual contributions.}
The classical (1,1,0) length change possesses two-behaviours in the Kagome ice phase, due to the degeneracy of the Kagome ice manifold being reflected in the length change Eq. \ref{eq_110_non_kramers_length}.
\textcolor{black}{The average over the degenerate branches matches up with the ED result for the (1,1,0) direction.}
The values of the chosen lattice-pseudospin couplings are presented in Supplementary Information V.}
\label{fig_NK_CSI_length_change_quad}
\end{figure*}
%%%

\section{Magnetostriction behaviours of Non-Kramers Pyrochlore Materials}

Under an applied magnetic field, 
the magnetic dipole moment couples at linear order and to the quadrupolar moments at quadratic order,
%%%
\begin{align}
\label{eq_mag_nk}
\mathcal{H}_{\text{mag,NK}} = & - {\bm{h}} \cdot \sum_t \sum_{\alpha = 0} ^{3} \hat{z}_{\alpha} \sfS _{t, (\alpha)} ^z \nonumber \\
& - \delta_1 \sum_t \sum_{\alpha = 0} ^{3} \left( h_\alpha^{x}  h_\alpha^{z} \sfS _{t, (\alpha)} ^x + h_\alpha^{y}  h_\alpha^{z} \sfS _{t, (\alpha)} ^y \right)  \\
& - \delta_2 \sum_t \sum_{\alpha = 0} ^{3} \left( \left[(h_\alpha^{y})^2  - (h_\alpha^{x})^2 \right] \sfS _{t, (\alpha)} ^x + 2h_\alpha^{x}  h_\alpha^{y} \sfS _{t, (\alpha)} ^y \right) \nonumber
\end{align}
%%%
where $\sum_t$ sums over all up tetrahedra, $\sfS _{t, (\alpha)} ^\mu$ is the $\sfS ^\mu$ moment on sublattice $\alpha$ of tetrahedron $t$. 
The second term in Eq. \ref{eq_mag_nk} is a quadratic-in-$h$ coupling to the quadrupolar moments, which is perturbatively weak as compared to the magnetic field coupling to the dipole moments; we include small phenomenological quantities $\delta_{12,}$ to represent this diminutive nature. {The strength of the quadratic background in the length change depends on $\delta_{1,2}$}.
\textcolor{black}{Indeed, the value of $\delta_{1,2}$ depends inversely on the gap ($\Delta$) between the ground state and excited states, $\delta_{1,2} \sim 1/ \Delta$.
In the context of Pr$_2$Zr$_2$O$_7$, the gap is relatively large ($\Delta \approx 9.5$meV \cite{kimura_nakatsuji_ncomm_pzo_2013}) thus physically justifying the minuscule magnitude of $\delta_{1,2}$. In other pyrochlore materials, such as Tb$_2$Ti$_2$O$_7$, this gap is almost an order of magnitude smaller ($\Delta \approx 1.4$ meV \cite{yb_gap}), and consequently the quadratic background to the magnetostriction is more dominant \cite{yb_magneto, yb_magneto_2}.}
\textcolor{black}{Applying the strategy described in }\textit{Methods}, we derive the following length change expressions along the $\bm{\ell}=(1,1,1)$ and ${\bm{\ell}} = (1,1,0)$ directions under ${\bm{h}} = \frac{h}{\sqrt{3}}(1,1,1)$,
\begin{widetext}
\begin{align}
\left(\frac{\Delta L }{L}\right)_{(1,1,1), {\rm NK}} ^{[111]} %= & \frac{\epsilon_B}{3}    +  \frac{2 \left(  \epsilon_{xy}   + \epsilon_{yz}  + \epsilon_{xz} \right)}{3} \nonumber \\
= & \frac{4}{3 c_{44}}  \left(2 k_1+k_2\right) \Bigg[\frac{1}{ \sqrt{3}} \left( 2\sfS_{(1)}^{x}- \sfS_{(2)}^{x}- \sfS_{(3)}^{x} \right)+ \left(\sfS_{(2)}^y- \sfS_{(3)}^y \right) \Bigg] + \frac{\left(g_3+g_4\right)}{3 \sqrt{3} c_B} h \Bigg[3 \sfS_{ (0) }^z-\sfS_{ (1) }^z-\sfS_{ (2) }^z-\sfS_{ (3) }^z \Bigg] \nonumber \\
& - \frac{2 \sqrt{3} }{27 c_{44}} h \Bigg[ \left(3g_3-6 g_4\right) \left(9 \sfS_{(0)}^z+\sfS_{(1)}^z+\sfS_{(2)}^z+\sfS_{(3)}^z \right)+ (32 g_1 + 16 g_2) \left(\sfS_{(1)}^z+\sfS_{(2)}^z+\sfS_{(3)} ^z \right) \Bigg],  \label{eq_111_non_kramers_length} \\ \nonumber \\
\left(\frac{\Delta L }{L}\right)_{(1,1,0), {\rm NK}} ^{[111]} %= & \frac{\epsilon_B}{3}    +  \frac{2 \left(  \epsilon_{xy}   + \epsilon_{yz}  + \epsilon_{xz} \right)}{3} \nonumber \\
= & - \frac{\left( 2k_1 + k_2\right)}{c_{44}}   \Bigg[\frac{1}{ \sqrt{3}} \left( \sfS_{(0)}^x - \sfS_{(1)}^x- \sfS_{(2)}^x + \sfS_{(3)}^x \right)+ \left(\sfS_{(0)}^y - \sfS_{(1)}^y - \sfS_{(2)}^y + \sfS_{(3)}^y \right) \Bigg] \nonumber \\
& +  \frac{\left( k_1 - k_2\right)}{4 (c_{11} - c_{12})}   \Bigg[\frac{1}{ \sqrt{3}} \left( \sfS_{(0)}^x + \sfS_{(1)}^x + \sfS_{(2)}^x + \sfS_{(3)}^x \right)+ \left(\sfS_{(0)}^y  + \sfS_{(1)}^y  + \sfS_{(2)}^y + \sfS_{(3)}^y \right) \Bigg]  \nonumber \\
& + \frac{\left(g_3+g_4\right)}{3 \sqrt{3} c_B} h \Bigg[3 \sfS_{ (0) }^z-\sfS_{ (1) }^z-\sfS_{ (2) }^z-\sfS_{ (3) }^z \Bigg] - \frac{ \sqrt{3} (g_1 - g_2) }{9 (c_{11} - c_{12})} h \Bigg[ \sfS_{(1)}^z+\sfS_{(2)}^z-2\sfS_{(3)}^z \Bigg] \label{eq_110_non_kramers_length} \\
& - \frac{ \sqrt{3} }{9 c_{44}} h \Bigg[ \left(3g_3-6 g_4\right) \left(3 \sfS_{(0)}^z+\sfS_{(1)}^z+\sfS_{(2)}^z-\sfS_{(3)}^z \right)+ (8 g_1 + 4 g_2) \left(\sfS_{(1)}^z+\sfS_{(2)}^z+2\sfS_{(3)}^z  \right) \Bigg] \nonumber,  
\end{align}
\end{widetext}
%%% 
%
where the subscript and superscript in $\left(\frac{\Delta L }{L}\right)$ refers to the magnetic field and length change directions, respectively. 
\textcolor{black}{We note that the pseudospin operators are taken to be understood as their expectation value with respect to the ground state.}
% $ S^j \rightarrow \langle S^j \rangle$ to clarify this point.
We note that we have redefined the couplings in Eqs. \ref{eq_111_non_kramers_length}, \ref{eq_110_non_kramers_length} from Eqs. \ref{eq_coupling_XY_NK}, \ref{eq_coupling_Z} for brevity i.e. $g_1 \equiv \frac{g_1}{\sqrt{6}}$, $g_2 \equiv \frac{g_2}{2\sqrt{3}}$, $g_3 \equiv \frac{2g_3}{3\sqrt{3}}$, and $g_4 \equiv \frac{g_4}{3\sqrt{3}}$, and $k_1 \equiv \frac{k_1}{\sqrt{3}}$, $k_2 \equiv \frac{k_2}{\sqrt{6}}$.
A striking observation of Eq. \ref{eq_111_non_kramers_length} is that uniform ferro-like ordering of the XY local moments results in vanishing length contributions from the quadrupolar moments. This scenario occurs for when $J_{\pm \pm} =0$ and only $J_{zz}, J_{\pm} > 0$.
\textcolor{black}{To have non-vanishing contributions from the quadrupolar moments, the ${\bm{\ell}} = (1,1,1)$ length change clearly requires the assistance of $J_{\pm \pm} \neq 0$ to give a non-uniformity (or even canting) to the ordering on each sublattice.}
For ${\bm{\ell}} = (1,1,0)$, a ferro-like ordering still yields a finite length change contribution from the quadrupolar moments.

\subsection{Unique magnetostriction signature of \textcolor{black}{multipolar} spin ice}

We present in Fig. \ref{fig_NK_CSI_length_change_quad} the unique magnetostriction behaviour for the MSI phase along the (1,1,1) and (1,1,0) direction under an applied field along the [111] direction.
The solid lines are obtained from a classical computation, while the squares are obtained from 32-site ED of the quantum model.
We describe in \textit{Methods} the procedure for these respective techniques.
The depicted shaded regions can be understood as a battle between two energy scales: spin-exchanges and magnetic field.
For small fields, the classical system enters into a Kagome ice (KI) phase, where the spin on sublattice 0 is fully polarized, while the remaining three spins conspire to satisfy the overall 2-in, 2-out ice rules of the exchange terms over an entire tetrahedron\cite{class_ki_1,class_ki_2,class_ki_3}.
\textcolor{black}{For the quantum model, the U(1) QSL survives for small window of magnetic field strength\cite{onada_qmc} (depicted by an amplified orange shaded region in Fig. \ref{fig_NK_CSI_length_change_quad}, for ease of viewing), until it enters into the quantum Kagome ice phase (described below).}
In the classical KI phase, the extensive degeneracy of the classical MSI phase partially remains within each Kagome layer\cite{class_ki_4,class_ki_5}.
This can be easily noticed on a given tetrahedron where three possible states satisfy the 2-in, 2-out ice rules: \textcolor{black}{$\{\sfS_{(0)}^{z}, \sfS_{(1)}^{z}, \sfS_{(2)}^{z}, \sfS_{(3)}^{z}  \}: \rm{(KI_a)}=\{\uparrow, \uparrow, \downarrow, \downarrow \}, \rm{(KI_b)} =\{\uparrow, \downarrow, \uparrow, \downarrow \}, \rm{(KI_c)}= \{\uparrow, \downarrow,\downarrow, \uparrow \}$.} 
\textcolor{black}{In the quantum limit, the ground state is that of a superposition over this degenerate manifold and is named quantum Kagome ice (QKI)}\cite{chm_qki,hermele_kagome_ice}.
Finally, in the large-field limit, all the pseudospins are polarized: this state corresponds to the pseudospins on sublattice-0 (sublattice-1,2,3) pointing out of (into) a given tetrahedron.
Interestingly, an island of XY quadrupolar ordering develops for intermediate fields during the transition between the KI and fully polarized state, where the XY orderings are not identical on the sublattices. 
\textcolor{black}{This results in a sharp discontinuity in the XY length change, accompanied by a `flip' in the z-component on sublattice 1 to $\sfS ^z <0$ (or any of the sublattices that have $\sfS ^z = 1/2$), and an alignment into the fully polarized state: $\{\sfS_{(0)}^{z}, \sfS_{(1)}^{z}, \sfS_{(2)}^{z}, \sfS_{(3)}^{z}  \} =\{\uparrow, \downarrow, \downarrow, \downarrow \} $.}
\textcolor{black}{We present in Supplementary Information VI the behaviour of the local pseudospin configurations on each sublattice under the [111] magnetic field.}

Both classical total length changes (the sum of the quadrupolar and dipolar contributions) in Fig. \ref{fig_NK_CSI_length_change_quad} possess a sharp discontinuity which originates from the transition from KI to the fully polarized phase.
Furthermore, both directions possess an additional quadratic-in-$h$ scaling behaviour (imposed on top of a linear-in-$h$ scaling from the dipole moments).
This arises from the $\sim h^2$ coupling of the magnetic field to the quadrupolar moments in Eq. \ref{eq_mag_nk}.
\textcolor{black}{A crucial difference between the two length change directions is mainly with the dipole contributions to the respective length change direction.
Firstly, in the fully polarized limit, (1,1,1) and (1,1,0) directions have opposite signs in their total length change. 
Secondly, and more interestingly, the (1,1,0) direction has two possible (dipole) length behaviours from the classical computation in the KI phase, while there is a unique behaviour for the (1,1,1) direction.
This is a result of the degeneracy of the KI phase.
For the (1,1,1) direction, all three degenerate KI states $\rm(KI_a,KI_b,KI_c)$ give the same expressions for the length change when these configurations are inserted into Eq. \ref{eq_111_non_kramers_length}.
For the (1,1,0) direction, however, the length change expression of Eq. \ref{eq_110_non_kramers_length} is sensitive to which of the three KI degenerate states is chosen.
In particular, $\rm(KI_a, KI_b)$ have the same length change behaviour when their respective configurations are inserted in to Eq. \ref{eq_110_non_kramers_length}, but $\rm(KI_c)$ has a different length change expression.
This difference can be traced to the dipole terms unique to the (1,1,0) direction in Eq. \ref{eq_110_non_kramers_length}. % that are responsible for the segregation of the KI's degenerate manifold.
Depending on which KI degenerate solution is chosen, we can thus get one of the two branches as depicted in Fig. \ref{fig_NK_CSI_length_change_quad}(b), and subsequently the sign of the dipole contribution for the (1,1,0) direction can flip (or be retained) in the fully polarized limit.
In a realistic system, it is possible that one may obtain an average over the three possible KI configurations.}
Interestingly, this `averaged' behaviour is precisely what 
the ED computation of the quantum model finds, corroborating the idea that the quantum ground state can be thought as the superposition of three degenerate configurations.
\textcolor{black}{All of these behaviours thus provide a sharp signature for the existence of a MSI phase at ${\bm{h}} = \bm{0}$.}

%%%
\begin{figure*}[t]
\includegraphics[width=0.6\linewidth]{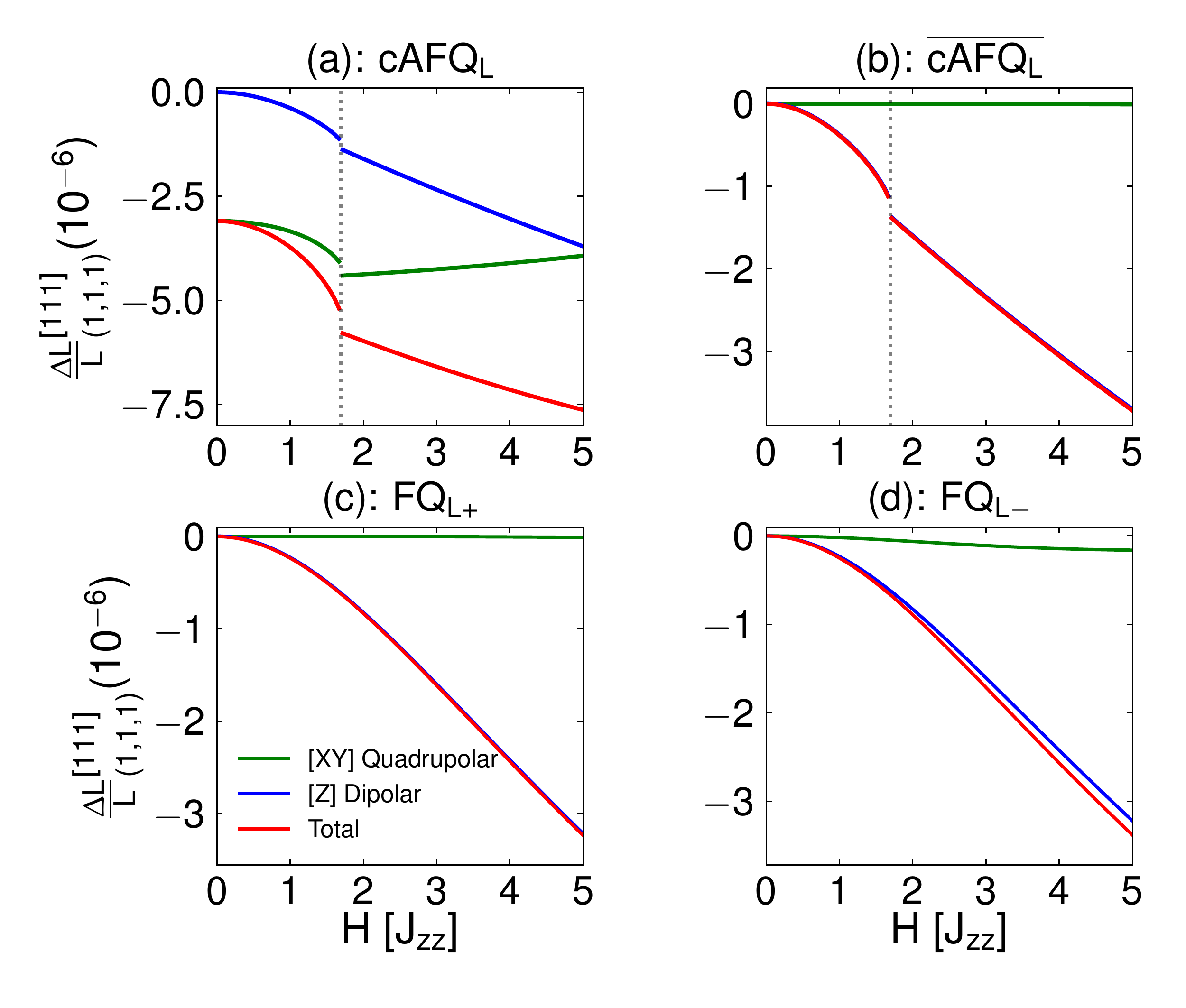}
\caption{ Length change, $\frac{\Delta L}{L}$, along the (1,1,1) direction under applied [111] magnetic field, ${\bm{h}}$, for the various classically multipolar ordered phases of non-Kramers ions ($J_{z\pm}=0$): (a) coplanar anti-ferroquadrupolar (cAFQ$\rm{_L}$), (b) a second coplanar anti-ferroquadrupolar ($\rm{\overline{cAFQ_L}}$), (c) and (d) ferro-quadrupolar ($\rm{FQ_{L\pm}}$) where the $\pm$ denote $J_{\pm \pm} > 0$ and $J_{\pm \pm} < 0$, respectively. The dashed vertical lines denote regions of discontinuity in the length change, intimately linked to the discontinuity arising from $\sfS _{(0)} ^z$ becoming fully polarized. 
The {green}, {blue} and {red} curves denote the length change arising from the XY pseudospin (quadrupolar), Z pseudospin (dipole), and combined contributions, respectively. 
\textcolor{black}{Just as in Fig. \ref{fig_NK_CSI_length_change_quad}, the dipole plot-line is (slightly) purposely shifted from the total length change plot-line to more easily visualize the individual contributions.}
The values of the chosen exchange couplings and lattice-pseudospin couplings are presented in Supplementary Information V.}
\label{fig_111_nk_others}
\end{figure*}
%%%

The ED results match well with the classical solution's (1,1,1) direction, especially in the region of KI and the fully polarized state.
Although the quadrupolar peak is broadened out as compared to its classical counterpart, we attribute this to \textcolor{black}{possible} finite size effects of ED, quantum fluctuations, and the challenge of extracting the symmetry-broken order parameter.  
We provide a detailed explanation of the latter point in \textit{Methods}.
For the (1,1,0) direction, more care needs to be taken to understand its apparent difference with its classical counterpart. 
In particular, the obtained ED ground state is non-degenerate, with the z-spin expectation value on each of sublattice-1,2,3 being $-\frac{1}{6}$.
This appears to suggest that the quantum ground state is an equal superposition over all the three degenerate classical Kagome ice states (on a given tetrahedron), resulting in the single possibility for the dipole length change. 
\textcolor{black}{For completeness, and to enable comparison with experiments, we present the length change under a [110] magnetic fields in Supplementary Information VII. Comparing the relatively smooth magnetostriction features for the [110] field to the peak structure for [111] field highlights the strong anisotropy and selection rules of magnetostriction.}

\subsection{Length change behaviours of non-Kramers multipolar ordered phases}
\label{sec_nk_mpos}

In order to clearly distinguish SI signatures from the other magnetically (or quadrupolar) ordered ground states, it is pertinent to consider the magnetostriction behaviour of the symmetry-broken ordered phases. 
\textcolor{black}{It is beneficial to first write, in terms of the classical multipolar order parameters (using Supplementary Information XIII), both the interacting pseudospin model,
%%%
\begin{align}
\label{eq_hamiltonian_classical_order_params_nk}
  {\mathcal{H}}^{\mathrm{tet}}_{\rm NK}=&
  \frac{1}{2} \Big[3J_{zz}m_{A_{2}}^2-6J_\pm {\bm{m}}_{E}^2+(2J_\pm-4J_{\pm\pm}) {\bm{m}}_{T_{1,B}}^2 \Big. \nonumber \\
  & \Big. -J_{zz} {\bm{m}}_{T_{1,A}}^2 +\left(2J_\pm+4 J_{\pm\pm}\right) {\bm{m}}_{T_{2}}^2 \Big]
\end{align}
%%%
and the $\bm{\ell} = (1,1,1)$ magnetostriction expression (Eq. \ref{eq_111_non_kramers_length}),
\begin{align}
\left(\frac{\Delta L }{L}\right)_{(1,1,1), {\rm NK}} ^{[111]} %= & \frac{\epsilon_B}{3}    +  \frac{2 \left(  \epsilon_{xy}   + \epsilon_{yz}  + \epsilon_{xz} \right)}{3} \nonumber \\
& =   \mathcal{Q}_0 \Big[m_{T_{2}}^x + m_{T_{2}}^y + m_{T_{2}}^z \Big]  - h \mathcal{C}_0 m_{A_{2}} \nonumber \\ 
& -h  (\mathcal{C}_1 + \mathcal{C}_2 ) \left(m_{T_{1,A}}^x + m_{T_{1,A}}^y + m_{T_{1,A}}^z \right)  \label{eq_111_non_kramers_length_order_params_main} %+  h (\mathcal{C}_1 + \mathcal{C}_2 ) \Big[m_{T_1}^x + m_{T_1}^y + m_{T_1}^z \Big]
\end{align}
%%%
where we drop the $g$erade subscript for the order parameters (\textcolor{black}{described in Fig. \ref{fig_classical_phases}}) for brevity, and collect the constants under $\mathcal{Q}_0 = \frac{8 \left(2 k_1+k_2\right)}{3 \sqrt{3} c_{44}} $, $\mathcal{D}_0 = \frac{4 \sqrt{3}\left(-8g_1+4g_2-3g_3+6g_4\right)}{9 c_{44}}$, $\mathcal{D}_1 = \frac{8 \sqrt{3} \left(  4g_1 + 2g_2 - 3g_3 + 6g_4 \right) }{27 c_{44}}$, and $\mathcal{D}_2 = \frac{2\left(g_3+g_4\right)}{3 \sqrt{3} c_B}$.}
Figure \ref{fig_111_nk_others} depicts the magnetostriction behaviour for ${\bm{\ell}} = (1,1,1)$ of the various multipolar ordered phases discussed earlier; the vertical dashed lines indicate jump discontinuous behaviours in the ordering.
\textcolor{black}{Specifically, both the cAFQ$\rm{_L}$ and $\rm{\overline{cAFQ_L}}$ have jump discontinuous behaviours which correspond to $\sfS _{(0)} ^z$ becoming fully polarized. The cAFQ$\rm{_L}$ also has the distinction of having a finite length change in the absence of an external field. \textcolor{black}{This is apparent from Eq. \ref{eq_111_non_kramers_length_order_params_main}, where the cAFQ$\rm{_L}$ order parameter is present even for $\bm{h} = \bm{0}$.}
The $\rm{FQ_{L\pm}}$ states do not possess any non-analytic behaviour in their length changes. Indeed, the local moments undergo a smooth and gradual change into the fully polarized state. 
We note that the $\rm{FQ_{L+}}$ and FQ$_{\rm L-}$ behaviours are related by a local $C_{4z}$ rotation of the pseudospins, where the $\pm$ denote $J_{\pm \pm} > 0$ and $J_{\pm \pm} < 0$, respectively.}

\textcolor{black}{Due to these mentioned characteristics, each of the MPOs have their own distinct signature that allows each of them to be identified individually, as well as be distinguished from MSI.
The $\rm{FQ_{L\pm}}$ states are the easiest to identify, as they possess a smooth change in the length change; this gradual change is not present in any of MSI nor the other MPOs.
cAFQ$\rm{_L}$ can also be distinguished as it holds the honour of being the only non-Kramers phase that has a finite length change in the absence of an external field. 
$\rm{\overline{cAFQ_L}}$ and MSI share some similarities, as both possess a jump discontinuity in the total length change.
However, a qualitative distinctions are (i) lack of a jump/peak in quadrupolar contribution for $\rm{\overline{cAFQ_L}}$, as compared to MSI, and (ii) the dipole length change contribution for $\rm{\overline{cAFQ_L}}$ phase \textcolor{black}{`drops below' (to a more negative length change) after the transition}, as compared to the MSI which jumps `above' (to a less negative length change). Furthermore, the MSI magnetostriction has a dominant linear-in-$h$ scaling behaviour for the MSI before the jump, while the $\rm{\overline{cAFQ_L}}$ has an overarching non-linear scaling before $\sfS _{(0)} ^z$ becomes fully polarized.
All of these differences demonstrate the uniqueness of the non-Kramers MSI and MPOs magnetostriction signatures.}

\section{Comparison with Kramers magnetically ordered phases}

\textcolor{black}
{A natural comparison is with the more prevalent Kramers ions of the pyrochlore family, which supports the usual magnetic dipole moments $J^{x,y,z}$.
Key differences between Kramers and non-Kramers ions are (i) $J_{z\pm}\neq 0$ which causes mixing between the spin ice and SFM phases, and (ii) the magnetic field can couple at linear order to the XY components of the pseudospins for Kramers ions.
These differences are a consequence of the pseudospin components being mere dipole moments and are thus odd under time-reversal.
We present the classical $T=0$ phase diagram and the magnetostriction behaviours of the magnetically ordered phases in Supplementary Information VIII, IX.
\textcolor{black}{The classical phase diagram for Kramers ions possess a variety of broken-symmetry phases: a generalized splayed ferromagnet (SFM), a coplanar antiferromagnetic Palmer-Chalker (PC) phase, and a 1D manifold of antiferromagnetic states.}
The magnetostriction behaviours for Kramers ions possess jump discontinuities in the length change for increasing field strengths, and for certain phases there are multiple such discontinuities.
}

\textcolor{black}{
We can draw specific contrasts between the associated Kramers and non-Kramers states.
For instance, the 1D manifold-like states in Kramers case have a discontinuity, while the corresponding $\rm{FQ_{L}}$ states of non-Kramers ions undergo smooth length change under increasing field.
Analogously, SFM and cAFQ$\rm{_L}$ can be distinguished as SFM has a vanishing length change at zero magnetic field, while for cAFQ$\rm{_L}$ it is finite.
Finally, PC and $\rm{\overline{cAFQ_L}}$ can be differentiated as PC has two discontinuous points in the length change, while $\rm{\overline{cAFQ_L}}$ has only one.
Such key differences in the length change behaviours of Kramers and non-Kramers ions highlight the broad applicability of magnetostriction in pyrochlore materials.
Furthermore, since Kramers ions are more commonly examined with conventional probes of magnetic ordering (most notably neutron scattering), magnetostriction can thus serve as useful corroborating evidence.
We contrast this with the non-Kramers situation, where there is a dearth of probes available, and where each of MPOs possess distinct features (Fig. \ref{fig_111_nk_others}) that allows each of them to be individually identified (and distinguished from non-Kramers MSI).
This comparison thus also serves to emphasize the suitability of magnetostriction to non-Kramers ions.
}

\section{Discussions}

In this work, we proposed that magnetostriction is an ideal probe of multipolar quantum spin ice ground states \textcolor{black}{and multipolar ordered states} in the pyrochlore oxide family. 
Employing a symmetry based approach, we constructed an elastic strain coupling to the local rare-earth moments, in the presence of an external magnetic field.
We studied all the possible classically ordered states of the pseudospin-1/2 model, and we also employed a 32-site exact diagonalization of the quantum model to examine the multipolar quantum spin ice in non-Kramers ions.
We found that the multipolar quantum spin ice phase in non-Kramers compounds has a unique magnetostriction behaviour, that is distinct from the other multipolar ordered phases \textcolor{black}{(which themselves have unique behaviours)}.

Our concrete theoretical results for magnetostriction under [111] and [110] magnetic fields provide a guide for targeted experimental investigations for MSI.
Indeed, recently presented experimental data of magnetostriction in Pr$_2$Zr$_2$O$_7$ \cite{nan_tang_aps_2019} seems to be qualitatively similar to our results.
Experimentally examining the length change for [110] magnetic field would be an important next step in verification of our selection rules of magnetostriction.
\textcolor{black}{Our study is also broadly applicable to other multipolar quantum spin candidate materials, Pr$_2$Sn$_2$O$_7$\cite{pso_3,pso_2,pso_1} and Pr$_2$Hf$_2$O$_7$ \cite{phfo_1,phfo_2,phfo_1,phfo_nat_phys}.}
In terms of future work, it would be interesting to examine finite temperature length change behaviours, such as thermal expansion.
Such studies would provide an insight into the non-trivial fractionalized excitations predicted in quantum spin ice, such as the emergent monopoles and photon.
It would also be intriguing to examine the study of magnetostriction in other frustrated lattices (with different symmetries) which are candidates for QSLs.
It would be fascinating to explore whether those systems also possess strong magnetostriction signatures, for both their proposed QSL and/or any nearby ordered phases.

\section{Methods}

\subsection{\textcolor{black}{Generalized magnetostriction expressions for non-Kramers ions}}

\textcolor{black}{To determine the magnetostriction (or total length change) expression ultimately requires knowledge of the relative length change in terms of the elastic strain components.
In that respect, we appeal to the relative length change expression of (as previously derived in Ref. \onlinecite{patri_unveil_2019}),
\begin{equation}
\left(\frac{\Delta L }{L}\right)_{\vec{\ell}} = \sum_{i,j=1} ^{3} \epsilon_{ij} \hat{\ell}_i \hat{\ell}_j \ , 
\label{eq_len}
\end{equation}
where $\epsilon_{ij} \equiv  \frac{1}{2}\left (\frac{\partial u_i}{\partial x_j} + \frac{\partial u_j}{\partial x_i} \right) $ is the standard strain tensor (in the global basis), and $\hat{\ell}_i$ is the $i^{th}$ component of the unit vector $\hat{\ell}$.
As is clear from Eq. \ref{eq_len}, determining $\epsilon_{ij}$ is essential to find the fractional length change.
To do so, the strain tensor in Eqs. \ref{eq_coupling_XY_NK}, \ref{eq_coupling_Z} first needs to rewritten in the global basis, using the change of basis described in Supplementary Information III.
This change of basis ensures that all elastic strain dependent quantities are written in terms of elastic strain normal modes in the global basis $\epsilon_{\mu, \nu, xy, xz, yz}$.
The subsequent total elastic free energy, $\mathcal{F}_{\text{elastic}} = \mathcal{F}_{\text{lattice}} + \mathcal{F}_{\text{XY,NK}} + \mathcal{F}_{\text{Z}}$, is then minimized with respect to the elastic strain normal modes to yield extremized elastic strain expressions, presented in Supplementary Information IV.
Finally, the extremized elastic strains are inserted into Eq. \ref{eq_len} to yield the generalized magnetostriction expressions along the direction of interest, $\bm{\ell}$.
}
\subsection{Classical solution to pseudospin-1/2 model}

The classical ground state of Eq. \ref{eq_hamiltonian_all} is that of a 4-sublattice ${\bm{q}} = {\bm{0}}$ ordering,
as Eq. \ref{eq_hamiltonian_all} can be written as the decoupled sum over individual (up or down) tetrahedron.
Thus any classical configuration that minimizes the energy of a single (up or down) tetrahedron automatically minimizes the total Hamiltonian.
The subsequent magnetic orderings over the entire pyrochlore lattice is that of magnetic orderings repeated over each tetrahedron. 
Since, the non-Kramers ions involve quadrupolar moments, these orderings are in fact multipolar orderings (rather than magnetic orderings as in the Kramers case).

%%%
\begin{figure*}[t]
\includegraphics[width=0.45\linewidth]{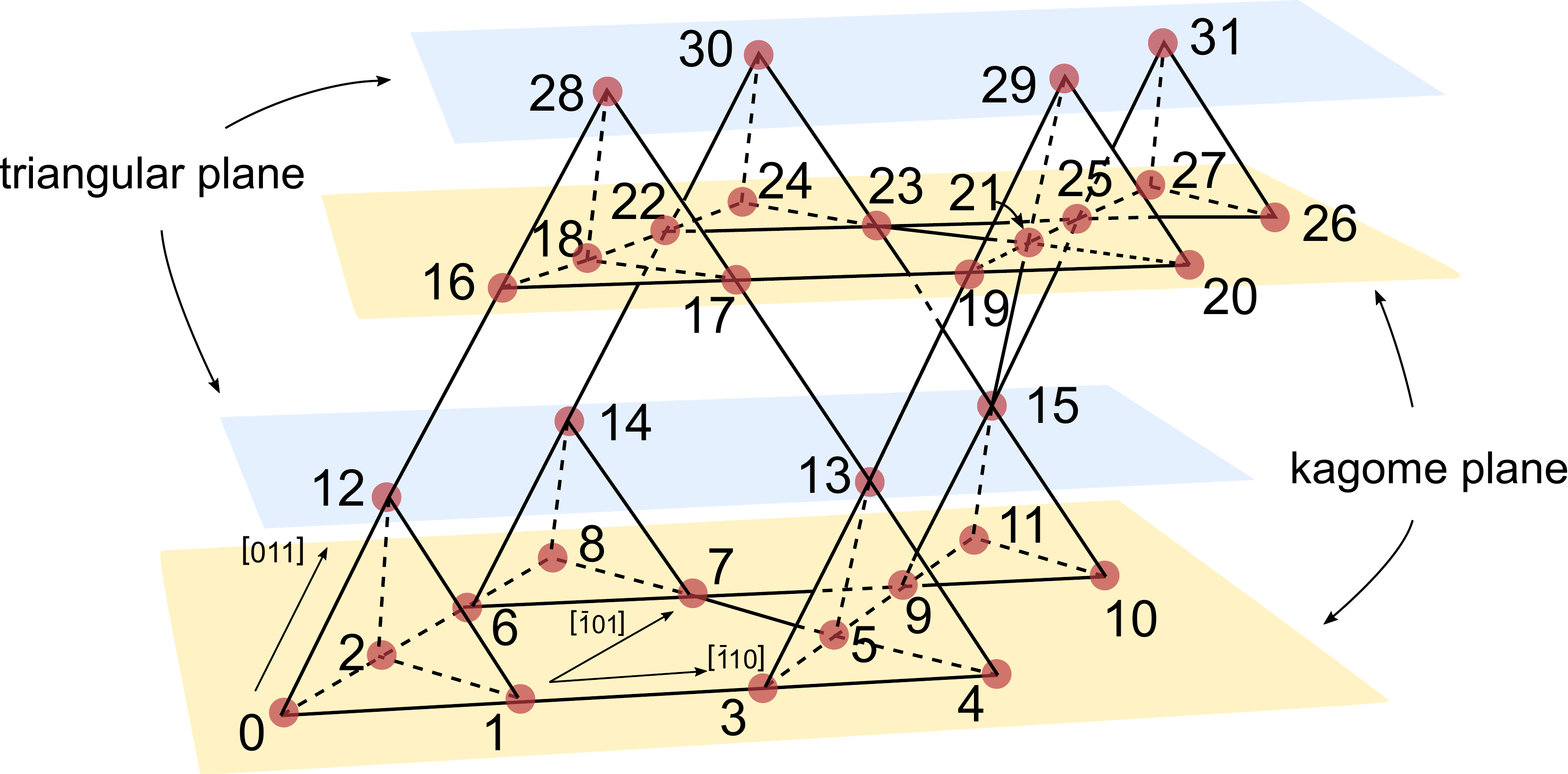} 
\includegraphics[width= 0.45\linewidth]{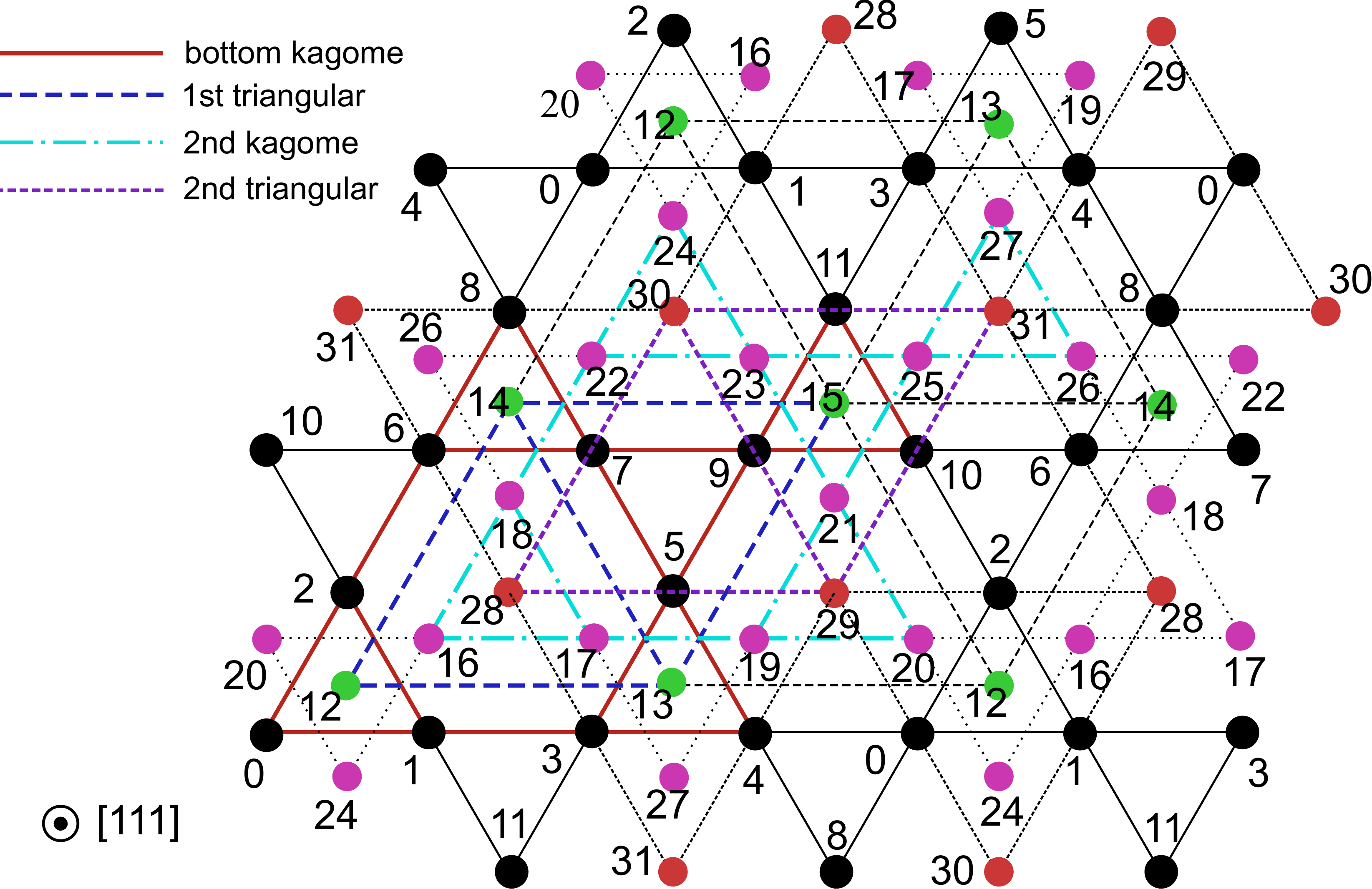} 
\caption{32 site ED cluster. 
Left Panel: schematic of the unit cell with the yellow (blue) planes denoting the Kagome (triangular) planes. 
Right Panel: top view of the cluster from the [111] direction. The four different colours denote sites on the four different triangular and Kagome layers in (a). The periodic directions are denoted by the axis.}
\label{fig_ed_32}
\end{figure*}
%%%

\subsection{32-site ED cluster study of quantum spin ice}

The ED ground state is obtained by employing the quantum lattice model solver package H$\Phi$\cite{h_phi_package}. 
A central step in this package is to represent the pseudospin operators in terms of fermionic operators, namely
%%%
\begin{equation}
\begin{aligned}
  \sfS_i^z&=\frac{1}{2}(c_{i,\uparrow}^\dagger c_{i,\uparrow}-c_{i,\downarrow}^\dagger c_{i,\downarrow}) \\
  \sfS_i^+&=c_{i,\uparrow}^\dagger c_{i,\downarrow}\\
  \sfS_i^-&=c_{i,\downarrow}^\dagger c_{i,\uparrow}.
\end{aligned}
\end{equation}
%%%
Using this formulation in Eq. \ref{eq_hamiltonian_all}, eigenenergies, eigenstates, one-body Green's function, and the two-body Green's function are obtained.
The one-body Green's function permits the extraction of the expectation value of the pseudospin operator i.e.  $\langle \sfS_i^\mu\rangle$.
The two-body Green's function allows the pseudospin-pseudospin correlation function to be obtained i.e. $\langle \sfS_i^\mu \sfS_j^\nu\rangle$.
The convergence factor of the Lanczos algorithm is determined by the condition of whether the relative error between the ground state energy at a given step and that of the previous step is less than $10^{-9}$.

We present in Fig. \ref{fig_ed_32}, the 32 site ED cluster we employ.
This cluster is formed by having two Bravais lattice points in each of the $x$, $y$, and $z$ directions; periodic boundary conditions are imposed in the three directions.
\textcolor{black}{Indeed, 32 site ED favourably compares to other numerical techniques such as numerical linked cluster methods in pyrochlore material Yb$_2$Ti$_2$O$_7$ \cite{changlani2017quantum}.}
In Supplementary Information XI, we present the computationally-less-intensive 16-site ED study from which we find the ${\bm{h}}={\bm{0}}$ phase diagram and associate the various phases with the ones present in Fig. \ref{fig_classical_phases}.
The location of the ED phase boundaries in the 16-site study guides us in choosing the $J_{\pm} = 0.02 J_{zz}$ and $J_{\pm \pm} = 0.05 J_{zz}$ parameter choice to investigate the quantum spin ice under a magnetic field.

An inherent challenge in ED studies is in extracting the pseudospin expectation values, as spontaneous symmetry breaking is only captured in the thermodynamic limit\cite{ed_order_parameter_2017}.
As such, in finite-sized clusters (in zero magnetic field), the pseudospin expectation values are always zero.
This issue can be avoided in a magnetic field, as the coupling of the field to the dipole moment explicitly breaks the symmetry, thus rendering a finite dipole expectation value.
However, in the absence of a field, we sketch the general strategy that can be used, where we employ the two-point pseudospin correlators and the intensity of the structure factor to divine the pseudo-spin expectation values.
In particular, we use the numerically produced $\langle \sfS_i ^z \rangle$ values as a benchmark to extract out the pseudospin expectation values of the $x$ and $y$ components.
Even though the quadrupolar contribution to the length change only involves the X and Y pseudospin components on sublattices-1,2,3, (as will become clear below), it is helpful to have information of the ordering on sublattice-0. 

We first recall that the pseudospin on sublattice 0 is fully polarized under a [111] field to yield a trivial ordering on the triangular layers of Fig. \ref{fig_ed_32}.
As such, we (reasonably) assume that we can decouple the two-point correlator for sublattice 0 sites ($i,j$) into the product of the expectation values on each site i.e. $\langle \sfS_i^\mu \sfS_j^\nu \rangle\simeq\langle \sfS_i^\mu\rangle\langle \sfS_j^\nu \rangle$ for  $i\neq j \in$ sublattice-0. Here $\mu = \{x,y,z\}$ are components of the pseduospin in the local axes.
This decoupling is true for the z-components (as it is trivially polarized), but we assume that we can also do so for the $x$ and $y$ components.
We now employ the intensity of the structure factor, which is defined as,
\begin{align}
  S^{\alpha\beta}({\bm q}) =\sum_\mu S^{\alpha\beta}_\mu({\bm q})=  \frac{1}{N_s}\sum_\mu\sum_{i\in\alpha,j\in\beta}e^{-i{\bm k}\cdot({\bm R}_i-{\bm R}_j)}\langle{  \sfS}_i^\mu  {  \sfS}_j^\mu \rangle,
\end{align}
where $\mu$ sums over the three components $\{x,y,z\}$ of the pseudospin, $\alpha$ and $\beta$ are sublattice indices $\{0,1,2,3\}$, $N_s$ is the total number of sites, $i,j$ are site locations of sublattice $\alpha, \beta$, respectively; in the $N_s=32$ site cluster, there are eight such $i,j$ locations each.
The wave number ${\bm k}$ is represented by using primitive reciprocal vectors ${\bm b}_i$ as
 ${\bm k}=\sum_{i=1}^3 q_i{\bm b}_i$. From this notation, we can easily notice that the first Brillouin zone is for $-1/2<q_i<1/2$.
Using this definition, we now extract the expectation value of the $\mu$ component of the pseudospin on sublattice-0 (hereafter represented as ``(0)"),
%%%
%\begin{equation}
  \begin{align}
    S_\mu^{00}(\bm {q =0})&=\frac{1}{N_s}\sum_{i,j\in(0)}\langle \sfS_i^\mu \sfS_j^\mu\rangle \label{eq_ed_32_s_mu_0} \\
    &\simeq \frac{1}{N_s}\left(\frac{N_s}{4}\langle (\sfS_{(0)}^\mu)^2\rangle+\frac{N_s}{4}\left(\frac{N_s}{4}-1 \right)\langle \sfS_{(0)}^\mu\rangle^2\right) \nonumber
  \end{align}
%\end{equation}
%%%
Here we used the fact that $\langle \sfS_i^\mu\rangle=\langle \sfS_{(0)}^\mu\rangle$ for $i\in$ (sublattce 0) due to the expected $\bm {q =0}$ ordering.
In Eq. \ref{eq_ed_32_s_mu_0}, the left-hand-side and the first term in the right-hand-side are numerically produced in the package, and so we solve for the second term.
We thus have an estimate of the pseudospin expectation value on sublattice-0; as a benchmark, the z-component expectation value from this method agrees to the numerically produced value to within 10$^{-5}$.

Confident in the above decoupling scheme, we now proceed to find the pseudospin expectation values on the other sublattices.
For sublattices-1,2,3 we consider two tetrahedra that are as far apart as possible in the 32-site cluster: tetrahedron A composed of site (12,0,1,2) and tetrahedron B composed of (31,25,26,27) in Fig. \ref{fig_ed_32};
here we label the sites on each tetrahedron in ascending order of the sublattice number i.e. $\{12,31\} \in(0)$, $\{0,25\} \in(1)$, $\{1, 26\} \in(2)$,  $\{2, 27\} \in(3)$.
Since these tetrahedra are well separated, we can consider the pseudospins on tetrahedron A to be to almost uncorrelated to the pseudospins on tetrahedron B, and thus assume they are separable just as we did above.
With this assumption, we then compute the expectation values for sublattices-1,2,3 using the following,

%%%
%\begin{widetext}
\begin{equation}
\begin{aligned}
  &\frac{\langle \sfS_{31}^\mu \sfS_{0}^\mu\rangle}{\langle \sfS_{(0)}^\mu\rangle} \simeq \frac{\langle \sfS_{31}^\mu\rangle\langle \sfS_{0}^\mu\rangle}{\langle \sfS_{(0)}^\mu\rangle}=
 \frac{ \langle \sfS_{(0)}^\mu\rangle\langle \sfS_{(1)}^\mu\rangle } {\langle \sfS_{(0)}^\mu\rangle}=\langle \sfS_{(1)}^\mu\rangle, \\
  & \frac{ \langle \sfS_{31}^\mu \sfS_{1}^\mu\rangle} {\langle \sfS_{(0)}^\mu\rangle} \simeq \frac{\langle \sfS_{31}^\mu\rangle\langle \sfS_{1}^\mu\rangle}{\langle \sfS_{(0)}^\mu\rangle}=
\frac{  \langle \sfS_{(0)}^\mu\rangle\langle \sfS_{(2)}^\mu\rangle}{ \langle \sfS_{(0)}^\mu\rangle}=\langle \sfS_{(2)}^\mu\rangle, \\
  &\frac{ \langle \sfS_{31}^\mu \sfS_{2}^\mu\rangle} {\langle \sfS_{(0)}^\mu\rangle} \simeq \frac{ \langle \sfS_{31}^\mu\rangle\langle \sfS_{2}^\mu\rangle } {\langle \sfS_{(0)}^\mu\rangle }=
  \frac{ \langle \sfS_{(0)}^\mu\rangle\langle \sfS_{(3)}^\mu\rangle } {\langle \sfS_{(0)}^\mu\rangle } =\langle \sfS_{(3)}^\mu\rangle .
\end{aligned}
\end{equation}
%\end{widetext}
%%%

Again as a benchmark, the pseudospin z-expectation values on the sublattices-1,2,3 computed from this method agree very well with the numerically produced value to within 10$^{-5}$, thus providing a validation for the assumption.
The above `correlator' strategy is beneficial when there is no explicit symmetry breaking field. 
Indeed, this is the case for the XY components when the perturbatively weak quadratic-in-$h$ coupling term is disregarded.
Comparing the `correlator' method's result for the XY expectation value with and without the quadratic-in-$h$ coupling term, we find that both results are the same to within $\sim 10^{-4}$. %, thus validating our result.
For completeness, the explicit expectation value (finite, in the presence of the quadratic-in-$h$ coupling term) gives the same qualitative peak feature (in the quadrupolar contribution to the length change) as the `correlator' method, except the quantitative value is reduced by a factor (as seen in Supplementary Information XII).

\section{Acknowledgments}
This work was supported by NSERC of Canada, and the Center for Quantum Materials at the University of Toronto. Y.B.K. is supported by the Killam Research Fellowship of the Canada Council for the Arts.
Computations were performed on the Niagara supercomputer at the SciNet HPC Consortium, and on the Graham supercomputer of Compute Canada. SciNet is funded by: the Canada Foundation for Innovation; the Government of Ontario; Ontario Research Fund - Research Excellence; and the University of Toronto.
\textcolor{black}{We thank Center for Advanced Computation at Korea Institute for Advanced Study for providing computing resources for this work.}
S.B.L. is supported by the National Research Foundation Grant No. NRF2017R1A2B4008097.
M.H. is supported by the Japan Society for the Promotion of Science through Program for Leading Graduate Schools (MERIT) and Overseas Challenge Program for Young Researchers.
We thank Satoru Nakatsuji, Mingxuan Fu, and Nan Tang for helpful
discussions on their experimental data on Pr$_2$Zr$_2$O$_7$ and sharing their
insights. After completing our work, we became aware of a parallel and
independent theoretical work by Subhro Bhattacharjee and Roderich Moessner,
concerning quantum spin ice physics of Pr$_2$Zr$_2$O$_7$. We thank them for
communicating their results to us.

\section{Author Contributions}

Y.B.K. conceived and supervised the research. A.S.P. and M.H. performed the calculations in this work. S.B.L. provided important preliminary results regarding the pseudospin-lattice couplings, and contributed to the discussion of the overall results. All authors contributed to writing the manuscript.
The authors declare no conflict of interest.

\clearpage

\beginsupplement

\section*{Supplementary Information}

\section{Local bases, psuedospin model matrices}
\label{app_local_bases}
The  pyrochlore lattice is an underlying face-centred cubic (FCC) Bravais lattice with four sublattices per unit cell.
We define the following local bases on each sublattice $\alpha$ in Table~\ref{tab_sublattice_basis}.
%%%
\begin{table}[h]
\centering
\begin{tabular}{c|c|c|c|c}
$\alpha$ & 0 & 1 & 2 & 3 \\
\hline 
$\hat{z}_{\alpha}$ & $\frac{1}{\sqrt{3}} \left(1,1,1 \right)$ & $\frac{1}{\sqrt{3}} \left(1,-1,-1 \right)$ & $ \frac{1}{\sqrt{3}} \left(-1,1,-1 \right)$ & $ \frac{1}{\sqrt{3}} \left(-1,-1,1 \right)$ \\
$\hat{x}_{\alpha}$ & $ \frac{1}{\sqrt{6}} \left(-2,1,1 \right)$ & $ \frac{1}{\sqrt{6}} \left(-2,-1,-1 \right)$ & $ \frac{1}{\sqrt{6}} \left(2,1,-1 \right)$ & $ \frac{1}{\sqrt{6}} \left(2,-1,1 \right)$ \\
$\hat{y}_{\alpha}$ &  $\frac{1}{\sqrt{2}} \left(0,-1,1 \right)$ & $\frac{1}{\sqrt{2}} \left(0,1,-1 \right)$ & $ \frac{1}{\sqrt{2}} \left(0,-1,-1 \right)$ & $ \frac{1}{\sqrt{2}} \left(0,1,1 \right)$
\end{tabular}
\caption{Local sublattice basis vectors.}
\label{tab_sublattice_basis}
\end{table}
%%%

Within the local bases, we employ the $\gamma_{ij}$ matrix which has the following matrix representation
\begin{align}
\gamma {=} \left(
\begin{array}{cccc}
 0 & 1 & w & w^2 \\
 1 & 0 & w^2 & w \\
 w & w^2 & 0 & 1 \\
 w^2 & w & 1 & 0 \\
\end{array}
\right)
\end{align}
where $w = e^{2 \pi i /3}$.

\section{Symmetry transformations of pseudospins, magnetic field, and elastic strain under $D_{3d}$}
\label{app_symm_local_global}

The $D_{3d}$ point group can be generated by the two following elements, which written in an orthonormal basis ($\mathbb{R}^3$ space) are,

\noindent\begin{minipage}{.5\linewidth}
\[
  \mathcal{S}_{6} ^{-}= 
\left(
\begin{array}{ccc}
 \frac{1}{2} & \frac{\sqrt{3}}{2} & 0 \\
 -\frac{\sqrt{3}}{2} & \frac{1}{2} & 0 \\
 0 & 0 & -1 \\
\end{array}
\right)  
\]
\end{minipage}%
\begin{minipage}{.5\linewidth}
\[
  \mathcal{C}_{21}^{'} = 
\left(
\begin{array}{ccc}
 -1 & 0 & 0 \\
 0 & 1 & 0 \\
 0 & 0 & -1 \\
\end{array}
\right)
\]
\end{minipage} \\
where $  \mathcal{S}_{6} ^{-}$ is an improper rotation about the z-axis by $\pi/3$, and $  \mathcal{C}_{21}^{'}  $ is a $\pi$ rotation about the y-axis. 
Using these generators, we can transform the pseudospin-1/2 quantities (on sublattice $\alpha$) as,
\noindent\begin{minipage}{.5\linewidth}
\begin{align}
&\sfS_\alpha ^x   \xrightarrow{\mathcal{S}_{6} ^{-}} -\frac{1}{2} S_x^{\alpha} - \frac{\sqrt{3}}{2} S_y^{\alpha} \nonumber \\
&\sfS_\alpha ^y  \xrightarrow{\mathcal{S}_{6} ^{-}} \frac{\sqrt{3}}{2} S_x^{\alpha} - \frac{1}{2}S_y^{\alpha} \nonumber \\
&\sfS_\alpha ^z  \xrightarrow{\mathcal{S}_{6} ^{-}} S_z^{\alpha} \nonumber
\end{align}
\end{minipage}%
\begin{minipage}{.5\linewidth}
\begin{align}
&\sfS_\alpha ^x   \xrightarrow{\mathcal{C}_{21}^{'}} S_x^{\alpha} \nonumber \\
&\sfS_\alpha ^y  \xrightarrow{\mathcal{C}_{21}^{'}} -S_y^{\alpha}  \\
&\sfS_\alpha ^z  \xrightarrow{\mathcal{C}_{21}^{'}} - S_z^{\alpha} \nonumber
\end{align}
\end{minipage} \\

\noindent\begin{minipage}{.5\linewidth}
\begin{align}
&h_\alpha ^x   \xrightarrow{\mathcal{S}_{6} ^{-}} -\frac{1}{2} h_\alpha^{x} - \frac{\sqrt{3}}{2} h_\alpha^{y} \nonumber \\
&h_\alpha ^y  \xrightarrow{\mathcal{S}_{6} ^{-}} \frac{\sqrt{3}}{2} h_\alpha^{x} - \frac{1}{2}h_\alpha^{y} \nonumber \\
&h_\alpha ^z  \xrightarrow{\mathcal{S}_{6} ^{-}} h_z^{\alpha} \nonumber 
\end{align}
\end{minipage}%
\begin{minipage}{.5\linewidth}
\begin{align}
&h_\alpha ^x   \xrightarrow{\mathcal{C}_{21}^{'}} -h_\alpha^{x} \nonumber \\
&h_\alpha ^y  \xrightarrow{\mathcal{C}_{21}^{'}} h_\alpha^{y}  \\
&h_\alpha ^z  \xrightarrow{\mathcal{C}_{21}^{'}} - h_\alpha^{z} \nonumber 
\end{align}
\end{minipage} \\
And finally, the elastic tensor transforms in the usual manner i.e. $ \overleftrightarrow{\epsilon} \rightarrow A \overleftrightarrow{\epsilon} A^{T} $, where $A$ is the symmetry element.

\section{Relating quantities in local axes to global axes}
\label{app_local_global_axes}
We present the transformation of the relevant quantities from the local axes to the global axes. The vector like quantities such as the magnetic field are transformed using ${\bm{h}}_{\alpha} = \mathbb{P}^{-1} _{\alpha} {\bm{h}}$, and the tensor-strain is transformed using ${\bm{\epsilon}}^{\alpha} = \mathbb{P}^{-1} _{\alpha} {\bm{\epsilon}} \ \mathbb{P} _{\alpha}$. Here $\mathbb{P}_{\alpha}$ is the change of basis matrix for sublattice-$\alpha$ i.e. its columns contain the basis vectors of the given subalttice as denoted in Table \ref{tab_sublattice_basis}. For concreteness, we present the local-to-global magnetic field transformations below,

\begin{align}
\left(
\begin{array}{c}
h^x _{(0)} \\
h^y _{(0)}\\
h^z _{(0)} \\
\end{array}
\right)
= 
\left(
\begin{array}{c}
 \frac{-2 h^x+h^y+h^z}{\sqrt{6}} \\
 \frac{h^z-h^y}{\sqrt{2}} \\
 \frac{h^x+h^y+h^z}{\sqrt{3}} \\
\end{array}
\right)
\end{align}

\begin{align}
\left(
\begin{array}{c}
h^x _{(1)} \\
h^y _{(1)}\\
h^z _{(1)} \\
\end{array}
\right)
= 
\left(
\begin{array}{c}
 -\frac{2 h^x+h^y+h^z}{\sqrt{6}} \\
 \frac{h^y-h^z}{\sqrt{2}} \\
 \frac{h^x-h^y-h^z}{\sqrt{3}} \\
\end{array}
\right)
\end{align}

\begin{align}
\left(
\begin{array}{c}
h^x _{(2)} \\
h^y _{(2)}\\
h^z _{(2)} \\
\end{array}
\right)
= 
\left(
\begin{array}{c}
 \frac{2 h^x+h^y-h^z}{\sqrt{6}} \\
 -\frac{h^y+h^z}{\sqrt{2}} \\
 -\frac{h^x-h^y+h^z}{\sqrt{3}} \\
\end{array}
\right)
\end{align}

\begin{align}
\left(
\begin{array}{c}
h^x _{(3)} \\
h^y _{(3)}\\
h^z _{(3)} \\
\end{array}
\right)
= 
\left(
\begin{array}{c}
 \frac{2 h^x-h^y+h^z}{\sqrt{6}} \\
 \frac{h^y+h^z}{\sqrt{2}} \\
 -\frac{h^x+h^y-h^z}{\sqrt{3}} \\
\end{array}
\right)
\end{align}
We reiterate that $h^{x,y,z}$ are magnetic field components in the global basis (i.e. no sublattice index).

\section{Extremized elastic strain tensor expressions for non-Kramers ions}
\label{app_extreme_strain_nk}

Extremizing the elastic free energy with respect to the normal modes ($\frac{\delta \mathcal{F}_{\text{elastic}}  } { \delta \epsilon_{ij} }=0$) yields the following expressions below.
We emphasize that the above expressions can be used (in combination with Eq. \ref{eq_len}) when finding the length change any direction of interest.
\begin{widetext}
\begin{align}
\epsilon_{\mu}^*  & =  \frac{g_1-g_2}{c_{11} - c_{12}} \Bigg[h^x \left(-\sfS_{(0)}^z-\sfS_{(1)}^z+\sfS_{(2)}^z+\sfS_{(3)}^z \right)+h^y \left(\sfS_{(0)}^z-\sfS_{(1)}^z+\sfS_{(2)}^z-\sfS_{(3)}^z\right)  \Bigg]   \nonumber \\ 
&  + \frac{k_1-k_2}{2 \left(c_{11} - c_{12} \right)} \Bigg[\sqrt{3} \left( \sfS_{(0)}^x+ \sfS_{(1)}^x+ \sfS_{(2)}^x+ \sfS_{(3)}^x \right)- \left(\sfS_{(0)}^y+\sfS_{(1)}^y+\sfS_{(2)}^y+\sfS_{(3)}^y \right)  \Bigg]  \\ \nonumber
\end{align}
% \frac{1}{6}
%
\begin{align}
\epsilon_{\nu}^* &=  \frac{g_1-g_2}{\sqrt{3} \left(c_{11} - c_{12} \right) } \Bigg[h^x \left(\sfS_{(0)}^z+\sfS_{(1)}^z-\sfS_{(2)}^z-\sfS_{(3)}^z\right)+h^y \left(\sfS_{(0)}^z-\sfS_{(1)}^z+\sfS_{(2)}^z-\sfS_{(3)}^z\right)  + 2 h^z \left(-\sfS_{(0)}^z+\sfS_{(1)}^z+\sfS_{(2)}^z-\sfS_{(3)}^z\right)\Bigg] \nonumber \\
&  - \frac{k_1-k_2}{2 \left(c_{11} - c_{12} \right)} \Bigg[ \left(\sfS_{(0)}^x+\sfS_{(1)}^x+\sfS_{(2)}^x+\sfS_{(3)}^x \right) +\sqrt{3} \left(\sfS_{(0)}^y+\sfS_{(1)}^y+\sfS_{(2)}^y+\sfS_{(3)}^y \right) \Bigg] 
\end{align}
\begin{align}
\epsilon_{B}^* &= \frac{g_3+g_4}{c_B} \Bigg[h^x \left(\sfS_{(0)}^z+\sfS_{(1)}^z-\sfS_{(2)}^z-\sfS_{(3)}^z \right)+h^y \left(\sfS_{(0)}^z-\sfS_{(1)}^z+\sfS_{(2)}^z-\sfS_{(3)}^z \right)+h^z \left(\sfS_{(0)}^z-\sfS_{(1)}^z-\sfS_{(2)}^z+\sfS_{(3)}^z \right)  \Bigg]
\end{align}

\begin{align}
\epsilon_{xy}^* &=  \frac{4 g_1+2 g_2-3 g_3+6 g_4 }{3 c_{44}} \Bigg[h^x \left(\sfS_{(0)}^z-\sfS_{(1)}^z+\sfS_{(2)}^z-\sfS_{(3)}^z \right)+h^y \left(\sfS_{(0)}^z+\sfS_{(1)}^z-\sfS_{(2)}^z-\sfS_{(3)}^z \right)\Bigg] \nonumber \\
& -  \frac{8 g_1+4 g_2+3 g_3-6 g_4}{3 c_{44}}  \Bigg[ h^z \left(\sfS_{(0)}^z+\sfS_{(1)}^z+\sfS_{(2)}^z+\sfS_{(3)}^z\right) \Bigg] \\
& - \frac{2 k_1+k_2}{c_{44}} \Bigg[\frac{1}{\sqrt{3}} \left( \sfS_{(0)}^x- \sfS_{(1)}^x- \sfS_{(2)}^x+ \sfS_{(3)}^x \right)   +  \left(\sfS_{(0)}^y-\sfS_{(1)}^y-\sfS_{(2)}^y+\sfS_{(3)}^y\right)\Bigg] \nonumber
\end{align}

\begin{align}
\epsilon_{xz}^* &=  \frac{4 g_1 + 2 g_2 - 3g_3 + 6 g_4 }{3 c_{44} } \Bigg[h^x \left(\sfS_{(0)}^z-\sfS_{(1)}^z-\sfS_{(2)}^z+\sfS_{(3)}^z  \right) +h^z \left(\sfS_{(0)}^z+\sfS_{(1)}^z-\sfS_{(2)}^z-\sfS_{(3)}^z  \right) \Bigg] \nonumber \\
& -  \frac{8 g_1+4 g_2+3 g_3-6 g_4}{3 c_{44} }  \Bigg[ h^y \left(\sfS_{(0)}^z+\sfS_{(1)}^z+\sfS_{(2)}^z+\sfS_{(3)}^z\right) \Bigg] \\
& -  \frac{2 k_1+k_2}{c_{44}} \Bigg[ \frac{1}{ \sqrt{3} } \left( \sfS_{(0)}^x-\sfS_{(1)}^x+ \sfS_{(2)}^x- \sfS_{(3)}^x \right)+ \left( -\sfS_{(0)}^y +  \sfS_{(1)}^y- \sfS_{(2)}^y+ \sfS_{(3)}^y \right) \Bigg]   \nonumber
\end{align}

\begin{align}
\epsilon_{yz}^* &= \frac{4 g_1+2 g_2-3 g_3+6 g_4 }{3 c_{44}} \Bigg[h^y \left(\sfS_{(0)}^z-\sfS_{(1)}^z-\sfS_{(2)}^z+\sfS_{(3)}^z \right)+h^z \left(\sfS_{(0)}^z-\sfS_{(1)}^z+\sfS_{(2)}^z-\sfS_{(3)}^z \right)\Bigg]  \\
& -  \frac{8 g_1+4 g_2+3 g_3-6 g_4}{3 c_{44}}  \Bigg[ h^x \left(\sfS_{(0)}^z+\sfS_{(1)}^z+\sfS_{(2)}^z+\sfS_{(3)}^z\right) \Bigg]  + \frac{4 k_1+2k_2}{c_{44}} \Bigg[ \frac{1}{\sqrt{3}} \left( \sfS_{(0)}^x+\sfS_{(1)}^x-\sfS_{(2)}^x-\sfS_{(3)}^x  \right)  \Bigg]  \nonumber
\end{align}

\end{widetext}
where we use the superscript to denote the extremized elastic strain.
We note that the magnetic field has also been re-written in terms of the global basis as described in Supplementary Information \ref{app_local_global_axes}.
%\vfill

\section{Numerical values of chosen coupling constants chosen}
\label{app_coupling_const_plots}
We take $J_{zz}=1$ in this study.
For the SI magnetostriction behaviours, we choose $J_{\pm}/J_{zz} = 0.02 $ and $J_{\pm \pm}/J_{zz} = 0.05 $.
For the cAFQ$\rm{_L}$ magnetostriction, we choose $J_{\pm}/J_{zz} = -0.5 $ and $J_{\pm \pm}/J_{zz} = -0.5 $.
For the $\rm{\overline{cAFQ_L}}$ magnetostriction, we choose $J_{\pm}/J_{zz} = -0.5 $ and $J_{\pm \pm}/J_{zz} = 0.5 $.
For the $\rm{FQ_{L+}}$ magnetostriction, we choose $J_{\pm}/J_{zz} = 0.72 $ and $J_{\pm \pm}/J_{zz} = 0.5 $.
For the $\rm{FQ_{L-}}$ magnetostriction, we choose $J_{\pm}/J_{zz} = 0.72 $ and $J_{\pm \pm}/J_{zz} = -0.5 $.
\textcolor{black}{We take $g_1 = g_2 = -\frac{9}{4\sqrt{3}} \times 10^{-7}$, $g_3 = 14\sqrt{3} \times 10^{-7}$, $g_4 = 4\sqrt{3} \times 10^{-7} $, $k_1 = -4.5\sqrt{3} \times 10^{-7}$, $k_2 = -2.6\sqrt{3} \times 10^{-7}$}, and $c_B = c_{44} = c_{11}-c_{22} = 1$.
Finally, we take $\delta_1 = 0.00075$ and $\delta_2 = -0.000088$ to emphasize the perturbative nature of the quadratic-in-$h$ magnetic field coupling.
%{The choice of the pseudospin-lattice coupling constants provides a length change scale that is comparable to experimental magnetostriction measurements in other Pr-based, heavy fermion, and Kitaev materials \cite{gegenwert_pr, ce_hfm, magneto_kitaev}.}
%
\textcolor{black}{The numerical values for the pseudospin-lattice couplings are taken with comparison to an experimental study of Pr-based heavy fermion compound, PrIr$_2$Zn$_{20}$ \cite{gegenwert_pr}.
PrIr$_2$Zn$_{20}$ shares similarities with Pr$_2$Zr$_2$O$_7$ in that both their interesting phenomena arise from Pr ions' $f^2$ electrons.
Taking the above coupling constants yields magnetostriction behaviours that are of the same scale as the reported study. 
Indeed, the physical scale of  $(\Delta L /L) \sim 10^{-6}$ for the relative length change is also observed in magnetostriction studies in other $f$ electron heavy fermion compounds \cite{ce_hfm, ce_hfm_2}, as well as in pressurized Kitaev materials \cite{magneto_kitaev}.
The actual value (or ratio) of the coupling constants can be determined by employing the proposed length change behaviours in conjunction with experimental measurements.
As an example, by subtracting off the leading linear-in-$h$ scaling behaviour in the experimental length change measurements for the [111] field and (1,1,1) direction allows the determination of $(2k_1 + k_2)$ in Eq. 7 and subsequently $(k_1 - k_2)$ from the (1,1,0) length change in Eq. 8, as we have numerically computed the pseudospin configurations.}

\section{Non-Kramers spin ice local pseudospin configuration in [111] magnetic field}
\label{app_nk_si_111_local_spin_config}

\begin{figure}[h]
\includegraphics[width=0.99\linewidth]{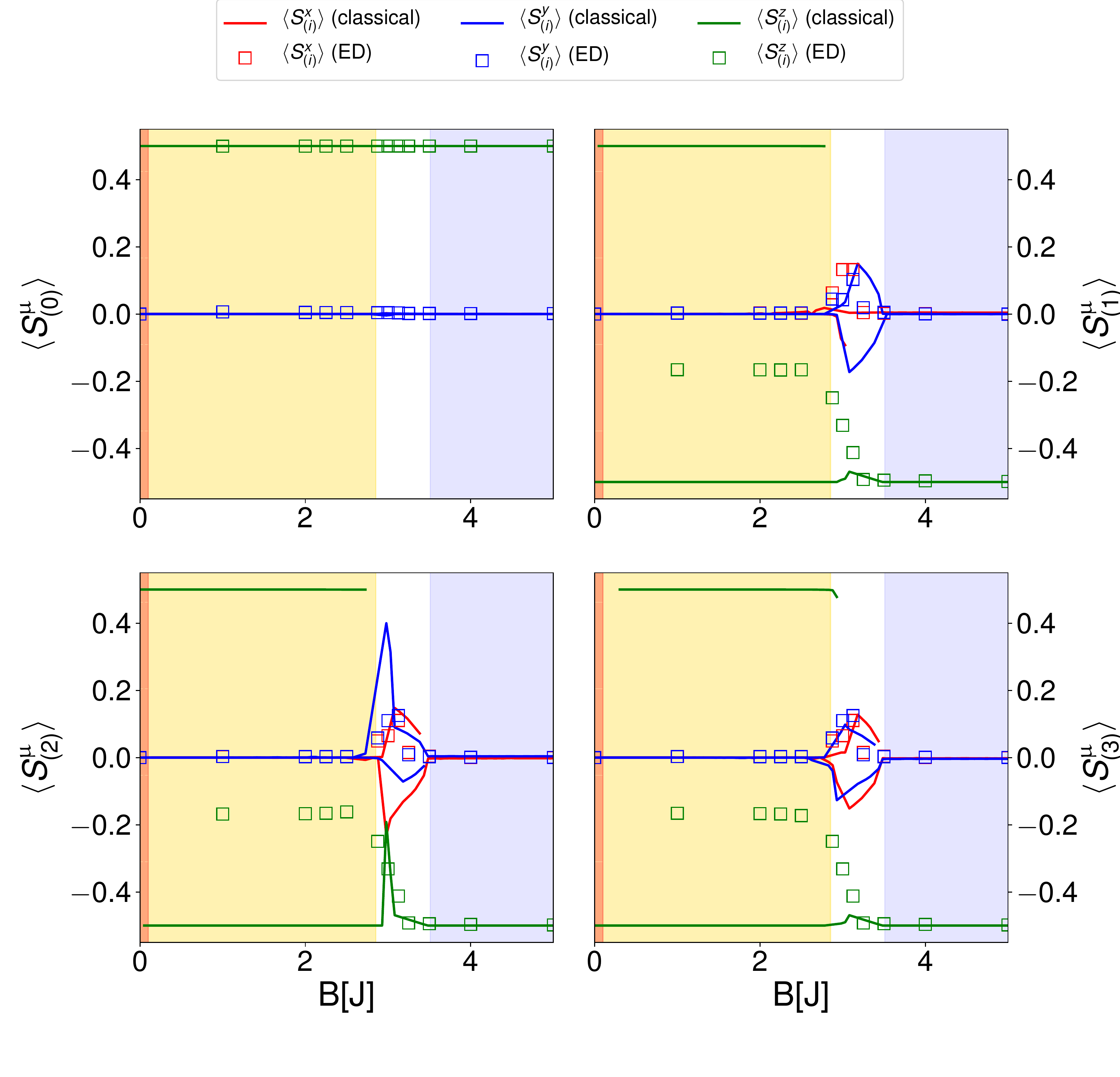}
\caption{ Local pseudospin configuration for non-Kramers SI in a [111] magnetic field. The degenerate Kagome spin ice phase is denoted by yellow shaded region, stable quantum spin ice is indicated by the orange shaded region, and the fully-polarized state depicted by indigo-blue shaded region.
Solid lines (unfilled squares) indicate classically (ED) obtained local pseudospin configurations on each sublattice.}
\label{fig_nk_si_order_params}
\end{figure}

We present in Fig. \ref{fig_nk_si_order_params} the behaviour of the local pseudospin configuration on each sublattice under the [111] magnetic field. 
The shaded regions in Fig. \ref{fig_nk_si_order_params} match up with the shaded regions in Fig. \ref{fig_NK_CSI_length_change_quad} of the main text: namely, `orange' region indicates the U(1) QSL (quantum spin ice),`yellow' region indicates Kagome Ice phase, and `indigo-blue' region indicates polarized phase.
The solid lines (unfilled squares) indicate the classically (`correlator' ED method) obtained local pseudospin configurations.
In the Kagome ice phase, the large degeneracy of the classical solution is clearly seen, while the ED study (as discussed in the main text) yields an averaged (over-all degenerate states) local pseudospin configuration.
%%%
%%%

%
\begin{figure*}[t]
\centering
\includegraphics[width=0.9\linewidth]{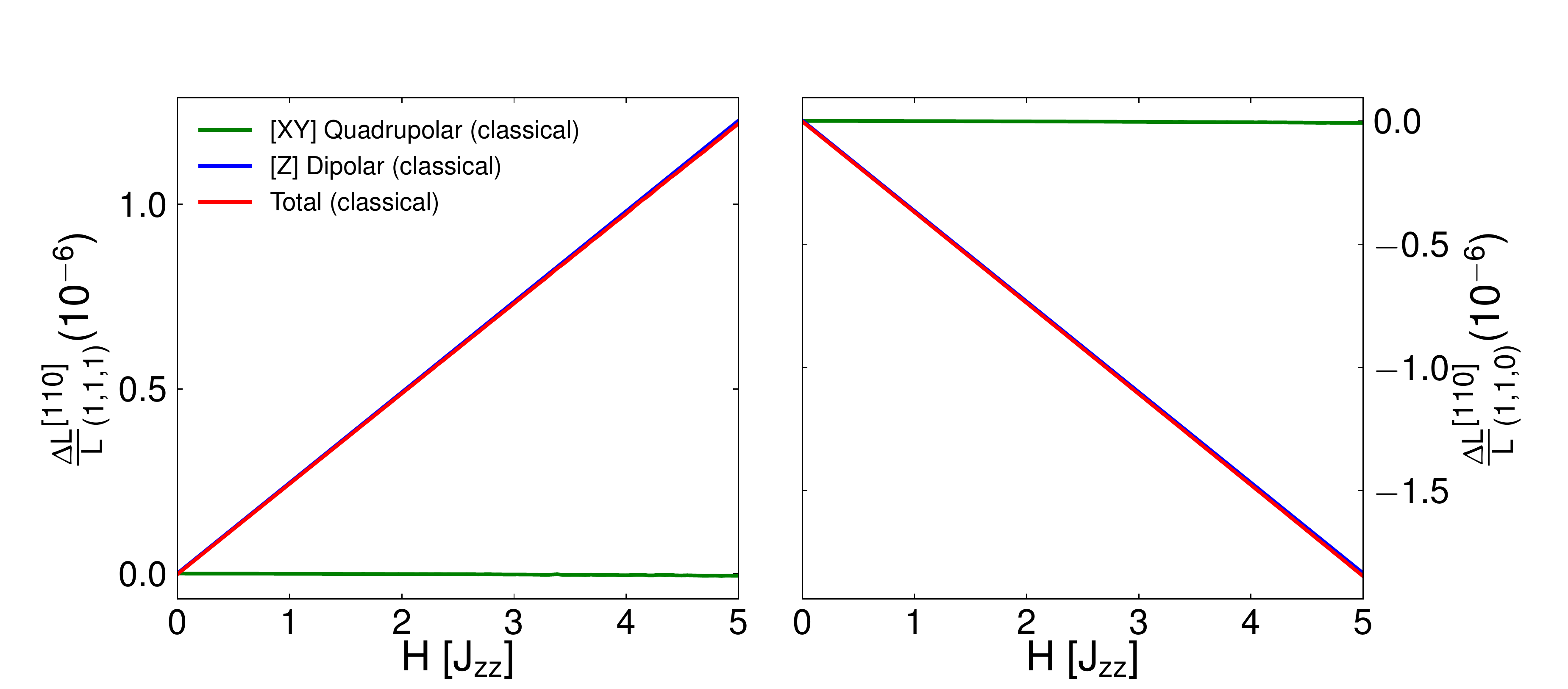}
\caption{ Length change, $\frac{\Delta L}{L}$, under applied [110] magnetic field, ${\bm{h}}$ for non-Kramers MSI phase $J_{\pm} = 0.02 J_{zz}, \  J_{\pm \pm} = 0.05 J_{zz}$. Left: along the (1,1,1) direction, Right: along the (1,1,0) direction.
Due to the lack of competition between the ice-rules and the magnetic field couplings, the length change has a monotonic behaviour.}
\label{fig_NK_CSI_110}
\end{figure*}

\section{Classical Non-Kramers Spin Ice in [110] magnetic field}
\label{app_nk_si_110}

We present in Fig. \ref{fig_NK_CSI_110} the classical magnetostriction behaviour along the (1,1,1) and (1,1,0) directions for a [110] magnetic field.
As seen, there is a lack of any clear/distinct features.
The reason lies with the fact that both the magnetic field and the ice rules can be simultaneously satisfied for this field direction.
For the [110] field, the magnetic field couples solely to sublattice-0,3 and fails to do so to sublattices 1,2 i.e. only $\hat{h} \cdot (\hat{x},\hat{y},\hat{z})_{0,3} \neq 0$, while $\hat{h} \cdot (\hat{x},\hat{y},\hat{z})_{1,2} = 0$.
As such, pseudospins on sublattice 0 and 3 respectively get aligned parallel ($+\hat{z}_0$) and anti-parallel ($-\hat{z}_3$) to the field, while the other sublattices conspire together to satisfy the ice rules i.e. partial degeneracy of the ice rules remains, with sublattices-1,2 taking $\sfS_{(1)}^z = \{\frac{1}{2}, -\frac{1}{2} \}$ and $\sfS_{(2)}^z = \{-\frac{1}{2}, \frac{1}{2}\}$, respectively.
This situation is valid for any finite field values.
Thus, there is no transition to any fully-polarized state in the large field limit, and no observable transition (unlike the [111] direction).

\section{Classical Phase diagram of Kramers ions}
\label{app_kramers_phase}

%%%
\begin{figure}[h]
\includegraphics[width=0.95\linewidth]{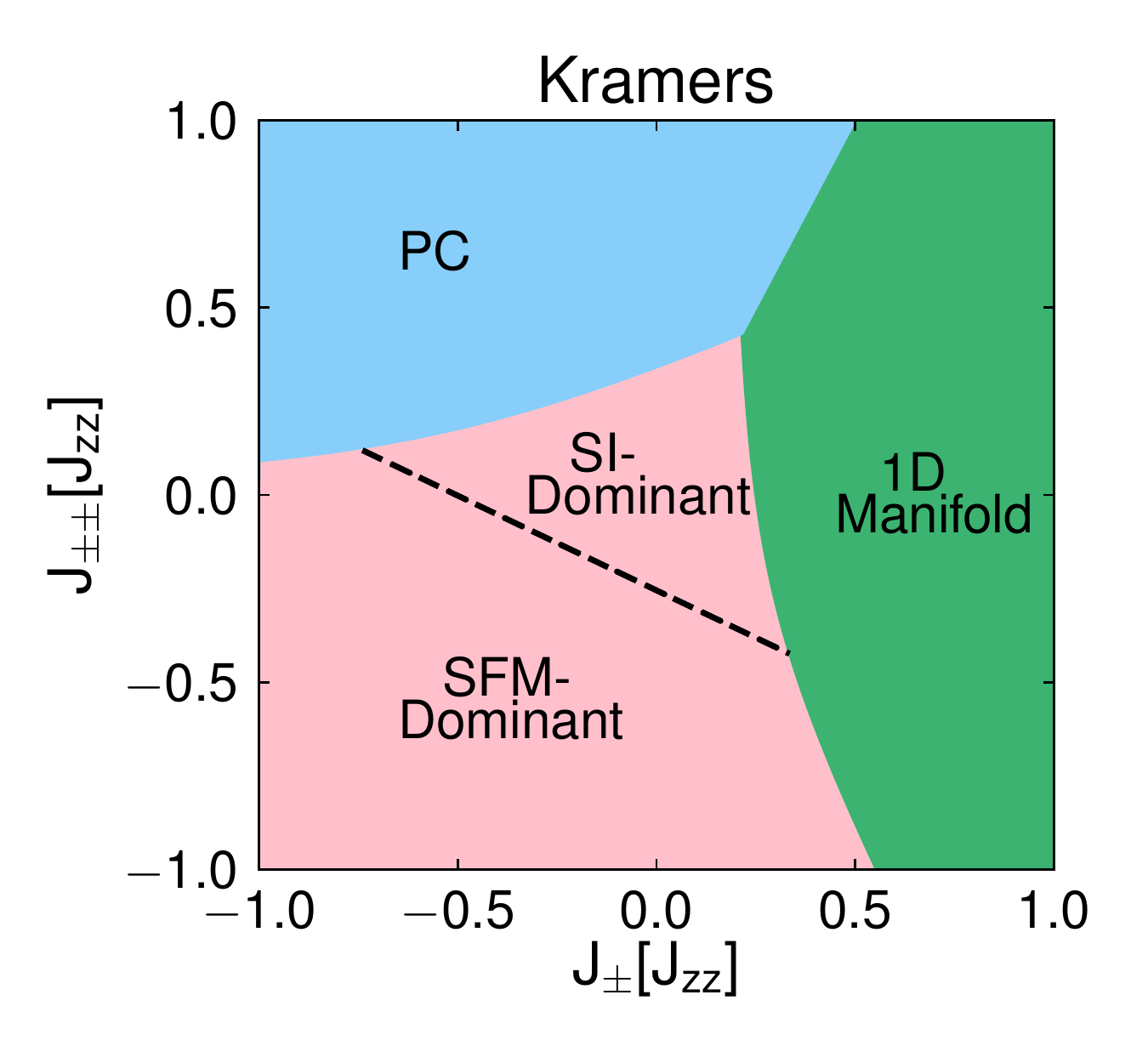}
\caption{ Classical Phase diagram of Eq. \ref{eq_hamiltonian_all} for Kramers ($J_{z\pm}=0.25 J_{zz}$). The depicted phases are spin ice (SI), splayed ferromagnet (SFM), Palmer-Chalker (PC), and 1D manifold of states.
The black dashed line is when the SI and SFM phases mix equally and separate a SI-Dominating phase and a SFM-dominating phase.}
\label{fig_classical_phases_kramers}
\end{figure}
%%%

The classical Kramers phase diagram of Fig. \ref{fig_classical_phases_kramers} provides a variety of possible phases: a blended phase composed of SI and splayed ferromagnet (SFM) of the same $T_{1g}$ symmetry, a coplanar antiferromagnetic Palmer-Chalker (PC) phase of $T_{2g}$ symmetry, and a 1D manifold of states with $E_g$ symmetry. 
An obvious distinction between the Kramers and non-Kramers phase diagram is that SI and SFM phases blend together for Kramers ions, while the corresponding non-Kramers phases are separated by a phase boundary for non-Kramers ions. This is a consequence of $J_{z\pm} \neq 0$, which allows the two aforementioned order parameters to mix.
One can easily notice this by expressing Eq. \ref{eq_hamiltonian_all} in terms of classical order parameters (orderings) on a single tetrahedron,

%%%
\begin{align}
\label{eq_hamiltonian_classical_order_params}
  {\mathcal{H}}^{\mathrm{tet}}=&
  \frac{1}{2} \Big[3J_{zz}m_{A_{2}}^2-6J_\pm {\bm{m}}_{E}^2+(2J_\pm-4J_{\pm\pm}) {\bm{m}}_{T_{2}}^2 \Big. \nonumber \\
  & \Big. -J_{zz} {\bm{m}}_{T_{1,A}}^2 +\left(2J_\pm+4 J_{\pm\pm}\right) {\bm{m}}_{T_{1,B}}^2 \Big.\\
   & \Big. -8J_{z\pm}  {\bm{m}}_{T_{1,A}}\cdot  {\bm{m}}_{T_{1,B}}\Big], \nonumber
\end{align}
%%%
where we use the definition of the order parameters as presented in Supplementary Information \ref{app_irreps}, and we drop the $g$erade subscript for the order parameters for brevity.
From the last term in Eq. \ref{eq_hamiltonian_classical_order_params} the two $T_{1}$ symmetry magnetic orderings mix with each other when $J_{z\pm} \neq 0$. Consequently, for Kramers ions, there exists a `common' region in the phase diagram with coexisting SI and SFM ordering. Depending on the location in this `common' region, the state is more SI-like or more SFM-like, which we label as SI-dominant and SFM-dominant, respectively. 

\vfill

\section{Magnetostriction Expressions of Kramers Pyrochlore Materials}
\label{app_kramers_all}
\begin{figure*}[t]
\includegraphics[width=0.8\linewidth]{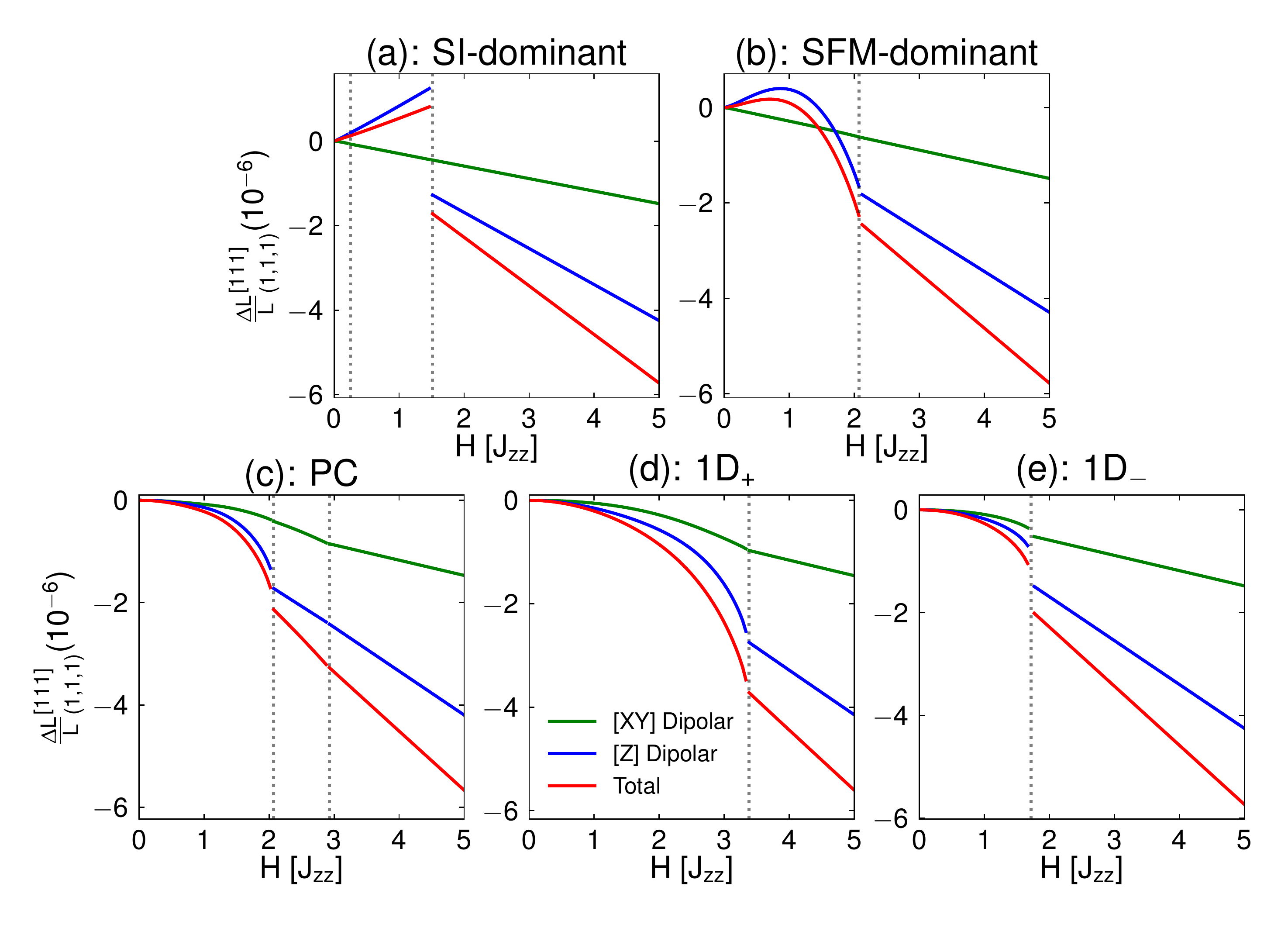}
\caption{ Length change, $\frac{\Delta L}{L}$, along the (1,1,1) direction under applied [111] magnetic field, $ {\bm{h}} $, for the various classically magnetically ordered phases of Kramers ions ($J_{z\pm}=0.25J_{zz}$). The dashed vertical lines denote regions of discontinuity in the length change, intimately linked to the discontinuity in the pseudospin expectation values. The {green}, {blue} and {red} curves denote the length change arising from the XY pseudospin (quadrupolar), Z pseudospin (dipole), and combined contributions, respectively. }
\label{fig_111_k_all}
\end{figure*}

We now turn to examining the magnetostriction behaviour of classically ordered Kramers phases.
Since all the pseudospin components are magnetic dipole moments, they all couple to an external magnetic field at linear order,
%%%
\begin{align}
\mathcal{H}_{\text{mag,K}} = -  {\bm{h}} \cdot \sum_t \sum_{\alpha = 0} ^{3} \Bigg[  \hat{z}_{\alpha} \sfS _{t, (\alpha)} ^{z}  + \frac{g_{xy}}{g_{zz}} \left(\hat{x}_{\alpha} \sfS _{t, (\alpha)} ^x +  \hat{y}_{\alpha} \sfS _{t, (\alpha)} ^y \right) \Bigg],
\end{align}
%%%
\noindent
where we include the non-vanishing $g$-tensor components $g_{xy}$, $g_{zz}$. As an archetypal example, we take the estimated $g$-tensor values of Yb$_2$Ti$_2$O$_7$\cite{Hodges_2001}: $(g_{xy}, g_{zz}) = (4.18, 1.77)$.
Due to the magnetic dipole nature of the XY moments, we have in addition to Eq. \ref{eq_coupling_Z}, 
%%%%
%\begin{widetext}
%\begin{align}
%\label{eq_coupling_XY_K}
%\mathcal{F}_{\text{XY,K}}  = & - n_0 \Big[ \sfS_\alpha ^{x} h_\alpha ^{x}  \epsilon_{xx}^{\alpha} + \sfS_\alpha ^{y} h_\alpha ^{y} \epsilon_{yy}^{\alpha} + \left(\sfS_\alpha ^{x} h_\alpha ^{y} + \sfS_\alpha ^{y} h_\alpha ^{x} \right) \epsilon_{xy}^{\alpha} \Big] \nonumber \\
%& - n_1 \Big[ \sfS_\alpha ^{x} h_\alpha ^{x}   \epsilon_{yy}^{\alpha} + \sfS_\alpha ^{y} h_\alpha ^{y} \epsilon_{xx}^{\alpha} - \left(\sfS_\alpha ^{x} h_\alpha ^{y} + \sfS_\alpha ^{y} h_\alpha ^{x} \right) \epsilon_{xy}^{\alpha}   \Big] \nonumber \\ 
%& -  n_2 \Big[ \left( \sfS_\alpha ^{y} h_\alpha ^{y} - \sfS_\alpha ^{x} h_\alpha ^{x} \right)  \epsilon_{xz}^{\alpha}  + \left( \sfS_\alpha ^{x} h_\alpha ^{y} + \sfS_\alpha ^{y} h_\alpha ^{x} \right) \epsilon_{yz}^{\alpha} \Big]  \nonumber \\ 
%& -  n_3 \Big[ \left( \sfS_\alpha ^{x} h_\alpha ^{x} + \sfS_\alpha ^{y} h_\alpha ^{y} \right)  \epsilon_{zz}^{\alpha}  \Big] -  n_5 \Big[  \sfS_\alpha ^{x}   \epsilon_{xz}^{\alpha} +  \sfS_\alpha ^{y}   \epsilon_{yz}^{\alpha}  \Big] h_\alpha ^{z} \nonumber \\
%%&  \nonumber \\
%& -  n_4 \Big[  \sfS_\alpha ^{x}   \left( \epsilon_{xx}^{\alpha}  - \epsilon_{yy}^{\alpha}  \right) -2  \sfS_\alpha ^{y}  \epsilon_{xy}^{\alpha} \Big] h_\alpha ^{z}   .
%\end{align}
%\end{widetext}
%%%%
%%%
\begin{widetext}
\begin{align}
\label{eq_coupling_XY_K}
\mathcal{F}_{\text{XY,K}}  = & - n_0 \Big[ \sfS_\alpha ^{x} h_\alpha ^{x}  \epsilon_{xx}^{\alpha} + \sfS_\alpha ^{y} h_\alpha ^{y} \epsilon_{yy}^{\alpha} + \left(\sfS_\alpha ^{x} h_\alpha ^{y} + \sfS_\alpha ^{y} h_\alpha ^{x} \right) \epsilon_{xy}^{\alpha} \Big]  - n_1 \Big[ \sfS_\alpha ^{x} h_\alpha ^{x}   \epsilon_{yy}^{\alpha} + \sfS_\alpha ^{y} h_\alpha ^{y} \epsilon_{xx}^{\alpha} - \left(\sfS_\alpha ^{x} h_\alpha ^{y} + \sfS_\alpha ^{y} h_\alpha ^{x} \right) \epsilon_{xy}^{\alpha}   \Big] \nonumber \\ 
& -  n_2 \Big[ \left( \sfS_\alpha ^{y} h_\alpha ^{y} - \sfS_\alpha ^{x} h_\alpha ^{x} \right)  \epsilon_{xz}^{\alpha}  + \left( \sfS_\alpha ^{x} h_\alpha ^{y} + \sfS_\alpha ^{y} h_\alpha ^{x} \right) \epsilon_{yz}^{\alpha} \Big]   -  n_3 \Big[ \left( \sfS_\alpha ^{x} h_\alpha ^{x} + \sfS_\alpha ^{y} h_\alpha ^{y} \right)  \epsilon_{zz}^{\alpha}  \Big] -  n_5 \Big[  \sfS_\alpha ^{x}   \epsilon_{xz}^{\alpha} +  \sfS_\alpha ^{y}   \epsilon_{yz}^{\alpha}  \Big] h_\alpha ^{z} \nonumber \\
%&  \nonumber \\
& -  n_4 \Big[  \sfS_\alpha ^{x}   \left( \epsilon_{xx}^{\alpha}  - \epsilon_{yy}^{\alpha}  \right) -2  \sfS_\alpha ^{y}  \epsilon_{xy}^{\alpha} \Big] h_\alpha ^{z}   .
\end{align}
\end{widetext}
%%%
where we again need the assistance of an external magnetic field in order to couple to the lattice strains, and we implicitly sum over $\alpha=\{0,1,2,3\}$. Employing the couplings in Eq. \ref{eq_coupling_Z} and Eq. \ref{eq_coupling_XY_K}, the parallel length change to a [111] magnetic field $ {\bm{h}} = \frac{h}{\sqrt{3}}(1,1,1)$ is
\begin{widetext}
\begin{align}
\label{eq_magnetostriction_K}
\left(\frac{\Delta L }{L}\right)_{(1,1,1), {\rm K}} ^{[111]} %= & \frac{\epsilon_B}{3}    +  \frac{2 \left(  \epsilon_{xy}   + \epsilon_{yz}  + \epsilon_{xz} \right)}{3} \nonumber \\
= & -h \Bigg[ \frac{1}{\sqrt{3}} \left(2 {\sfS ^x _{(1)}} -{\sfS^x_{(2)}}-{\sfS^x_{(3)}} \right)+ \left( {\sfS^y_{(2)}}- {\sfS^y_{(3)}} \right) \Bigg] \tilde{n} \nonumber \\
& +  \frac{\left(g_3+g_4\right)}{3 \sqrt{3} c_B} h \Bigg[3 \sfS^z_{ (0) }-\sfS_z^{ (1) }-\sfS^z_{ (2) }-\sfS^z_{ (3) } \Bigg] \\
& - \frac{2 \sqrt{3} }{27 c_{44}} h \Bigg[ \left(3g_3-6 g_4\right) \left(9 \sfS^z_{(0)}+\sfS^z_{(1)}+\sfS^z_{(2)}+\sfS^z_{(3)} \right)+ (32 g_1 + 16 g_2) \left(\sfS^z_{(1)}+\sfS^z_{(2)}+\sfS^z_{(3)}  \right) \Bigg] \nonumber
   \nonumber \\
& =  -\frac{2\tilde{n}}{\sqrt{3}} h \Big[m_{T_{1B}}^x + m_{T_{1B}}^y + m_{T_{1B}}^z \Big]  \label{eq_111_non_kramers_length_order_params} -h \Big[ \mathcal{D}_0 m_{A_{2g}} + (\mathcal{D}_1 + \mathcal{D}_2 ) \left(m_{T_{1A}}^x + m_{T_{1A}}^y + m_{T_{1A}}^z \right) \Big] \nonumber 
\end{align}
\end{widetext}
Due to the same form of the Z couplings as the non-Kramers case, the length change arising from the Z dipole moment is identical. The XY contribution has the same form (albeit accompanied by the magnetic field strength, $h$), with a complicated combination of elastic-pseudospin coupling constants, $\tilde{n} = \sqrt{3} \frac{ \left(2 c_B \left(5 \sqrt{2} {n_0}-3 \sqrt{2} {n_1}-4 {n_2}-2 \sqrt{2} {n_3}+4 {n_4}+\sqrt{2} {n_5}\right)+\sqrt{2} c_{44} ({n_0}+{n_1}+{n_3})\right)}{27 c_{44} c_B} $ is a collection of constants. 
We also collect the constants $\mathcal{D}_0 = \frac{4 \sqrt{3}\left(-8g_1+4g_2-3g_3+6g_4\right)}{9 c_{44}}$, $\mathcal{D}_1 = \frac{8 \sqrt{3} \left(  4g_1 + 2g_2 - 3g_3 + 6g_4 \right) }{27 c_{44}}$, and $\mathcal{D}_2 = \frac{2\left(g_3+g_4\right)}{3 \sqrt{3} c_B}$.
\textcolor{black}{We take $\tilde{n}=-1$ for the magnetostriction behaviours in Fig.} \ref{fig_111_k_all}, \textcolor{black}{and use the same choice of the other coupling constants as the non-Kramers case} (Supplementary Information \ref{app_coupling_const_plots}).

\section{Length change behaviours of Kramers magnetically ordered phases}

We present in Fig. \ref{fig_111_k_all} the magnetostriction behaviours of the magnetically ordered ground states. Due to the aforementioned mixing of the SI and SFM phases, we present the behaviour for choice of $J_\pm, J_{\pm \pm}, J_{z \pm}$ which yields dominant SI (SFM) behaviour over SFM (SI). Just as in Fig. \ref{fig_111_nk_others}, we denote jump discontinuous behaviours in the magnetic ordering by vertical dashed lines.

In all the Kramers ions behaviours, at $h=0$ the total length change vanishes as is apparent from Eq. \ref{eq_magnetostriction_K}.
Moreover, all phases possess a monotonically increasing XY contribution to the length change.
For the SI-dominant phase in Fig. \ref{fig_111_k_all}(a), there exist two points of discontinuity. The first (at small field) arises due to $\sfS _{(0)} ^z$ becoming fully polarized in the $-\hat{z}_0$ direction, and the second (at larger field) due to $\sfS _{(0)} ^z$ becoming polarized in the $+ \hat{z}_0$ direction. This discontinuity also appears (albeit less prominently) in the XY behaviour. 
The second discontinuity can be loosely associated to the discontinuity in the NK case, in that $\sfS _{(0)} ^z$ becomes fully polarized in both cases; however, since the magnetic field coupling involves $\sfS _{(0,1,2,3)}^{x,y}$, as well as the presence of the $J_{z \pm}$ term, it is not a direct comparison.
The SFM-dominant phase in Fig. \ref{fig_111_k_all}(b) possesses a single discontinuity, which (just as the second discontinuity point of the SI-dominant phase) is associated with $\sfS _{(0)} ^z$ becoming fully polarized. The broad maximum in the Z contribution arises due to a gradual change in the sign of $\sfS _{(0)} ^z$ from $\sfS _{(0)} ^z <0 $ in the SFM-like phase to the fully polarized $\sfS _{(0)} ^z = 1/2$.
The PC phase in Fig. \ref{fig_111_k_all}(c) also possesses two discontinuous points: the first associated with $\sfS _{(0)} ^z$ becoming fully polarized, and the second where $\sfS _{(1)} ^y \rightarrow 0$.
From the numerical minimization, the second discontinuity appears to be continuous.
Finally, the two 1D manifold states in Fig. \ref{fig_111_k_all}(d,e) have a single discontinuity again associated with $\sfS _{(0)} ^z$ becoming fully polarized.

As seen, there is a lack of clear difference between the various Kramers magnetically ordered phases.
In fact, only the SI-dominant phase appears to be distinct, with the dipole contribution flipping sign after the discontinuity.
This suggests that unlike the non-Kramers ions, magnetostriction is less suited for Kramers ions.

%%%
\begin{figure*}[t]
\includegraphics[width=0.43\linewidth]{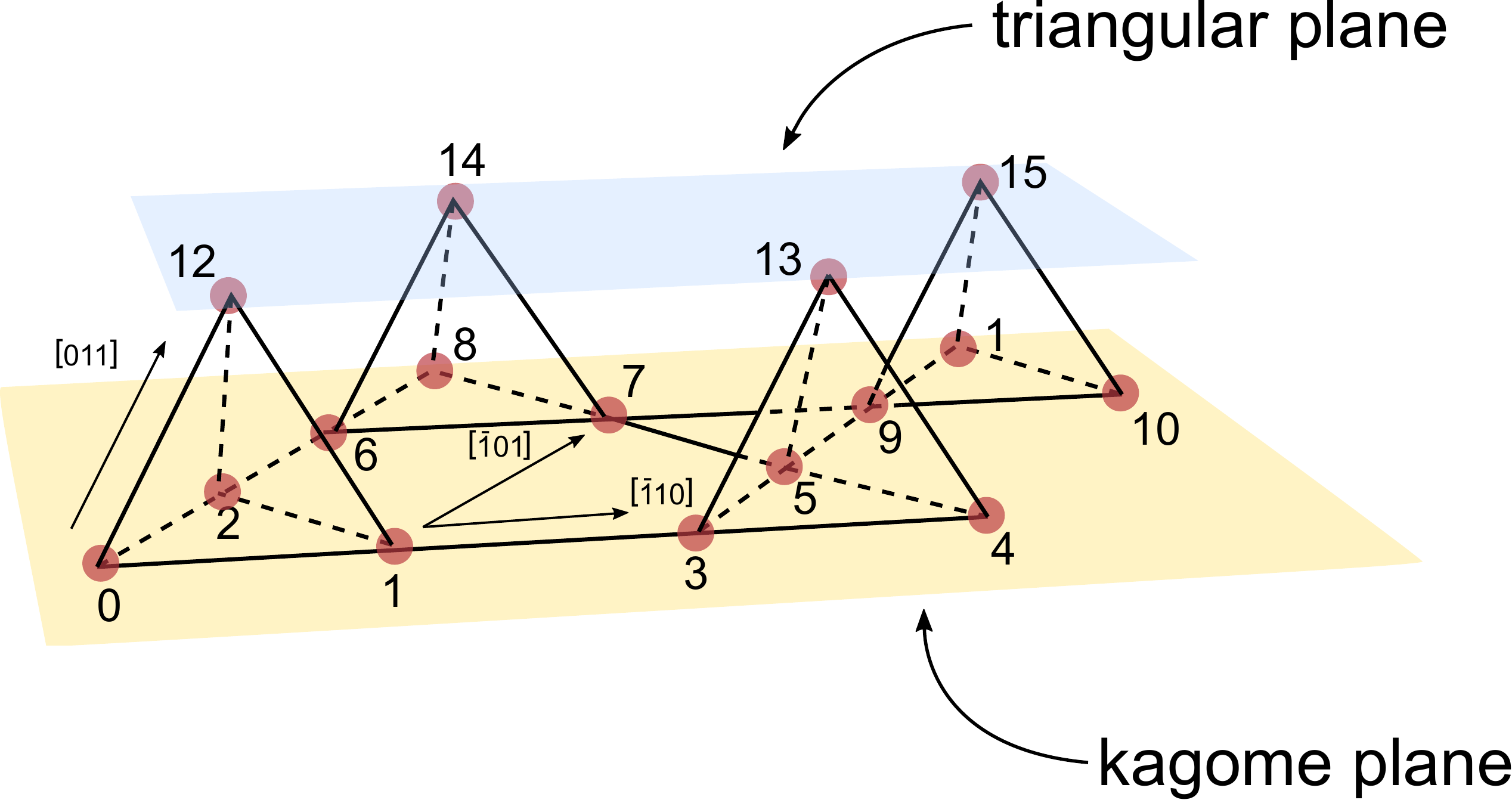} 
\includegraphics[width= 0.43\linewidth]{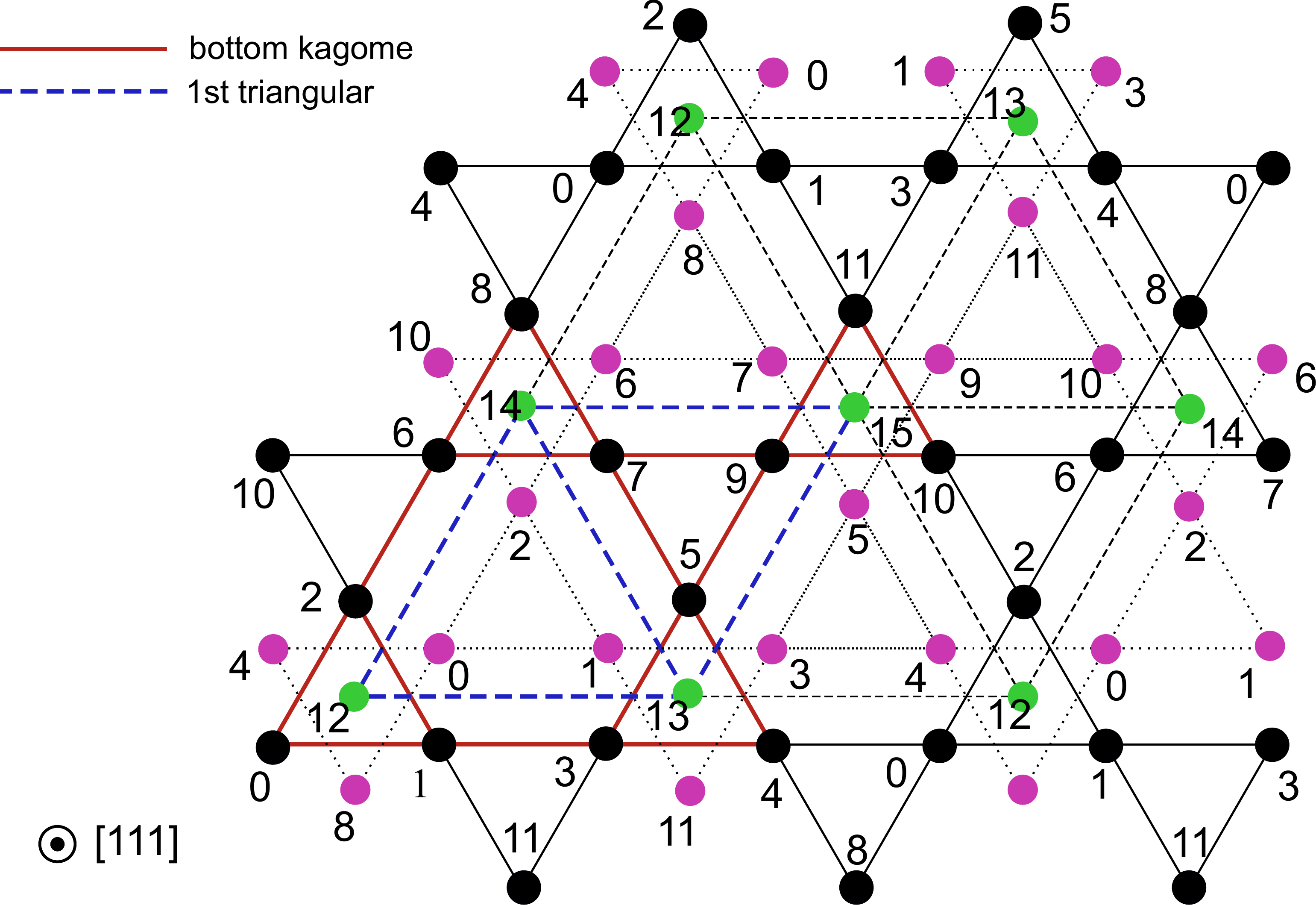} 
\caption{16 site ED cluster. 
Left Panel: schematic of the unit cell with the yellow (blue) planes denoting the Kagome (triangular) planes. 
Right Panel: top view of the cluster from the [111] direction. The four different colours denote sites on the four different triangular and Kagome layers in (a). The periodic directions are denoted by the axis.}
\label{fig_ed_16_schematic}
\end{figure*}
%%%

%\vfill

%\pagebreak

\section{16-site ED study of quantum spin ice}
\label{app_ed_16}

%%%
\begin{figure}[h]
\centering
\includegraphics[width=1\linewidth]{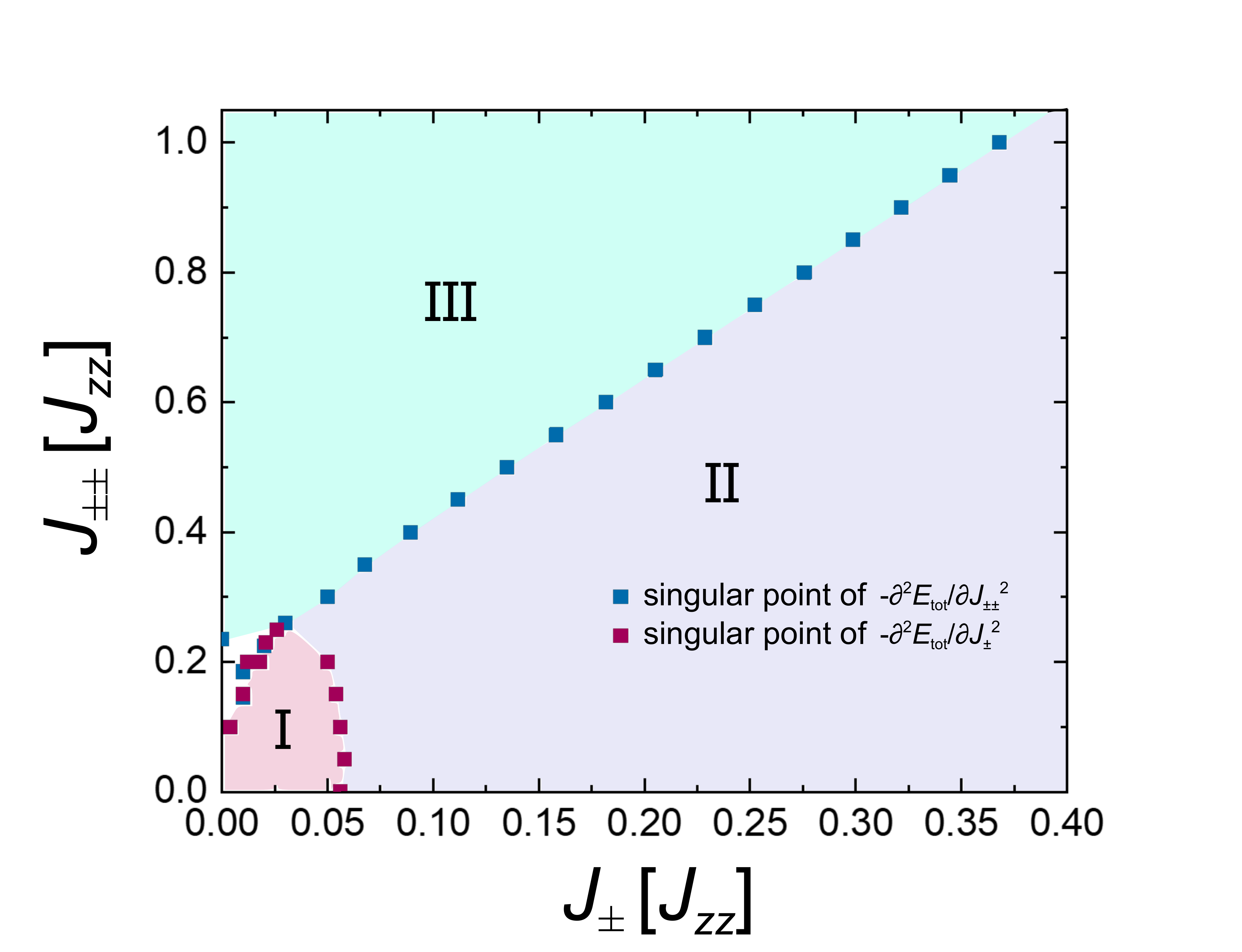}
\caption{ 16 site ED Phase diagram of Eq. \ref{eq_hamiltonian_all} for non-Kramers with $J_{\pm}=0.02J_{zz}, J_{\pm \pm} = 0.05 J_{zz}$.
The phase boundaries are denoted by the location of $\frac{\partial ^2 E}{\partial J_{\pm}^2}$ and $\frac{\partial ^2 E}{\partial J_{\pm \pm}^2 }$ becoming singular.}
\label{fig_ed_16_phases}
\end{figure}
%%%

%%%
\begin{figure*}[t!]
  \centering
  \includegraphics[width=\linewidth]{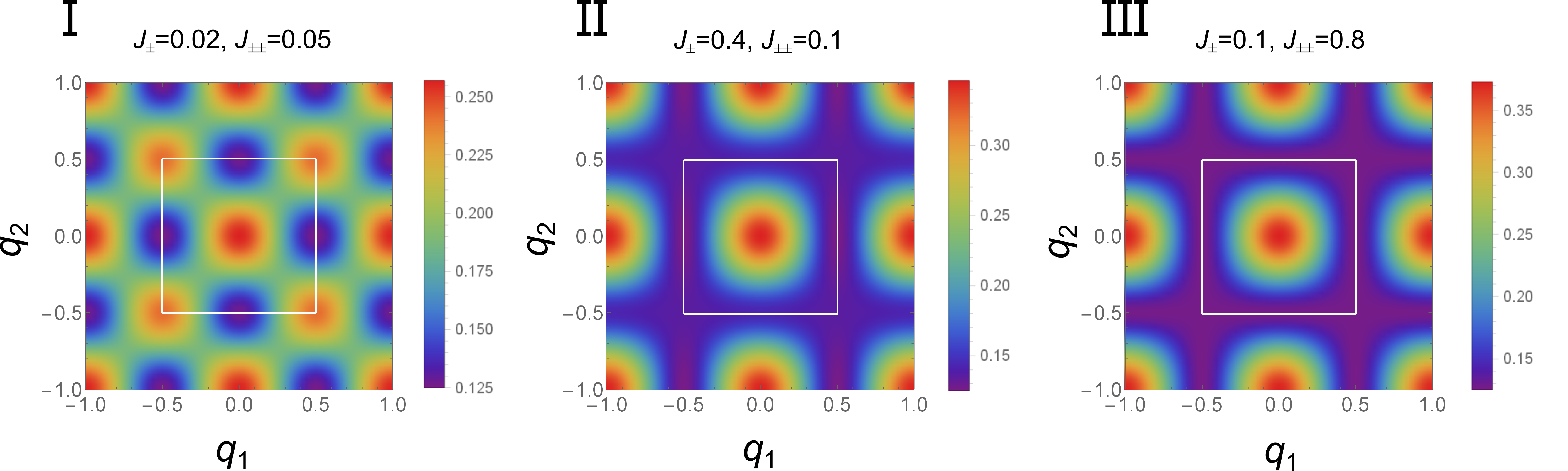}
  \caption{The intensity distribution of $S^{22}({\bm q})$ for each phases. The parameter sets are chosen as $(J_\pm,J_{\pm\pm})=(0.03,0.1)$
  for phase I, $(J_\pm,J_{\pm\pm})=(0.4,0.1)$ for phase II, and $(J_\pm,J_{\pm\pm})=(0.1,0.8)$ for phase III, respectively.}
  \label{fig_structure_factor_16}
\end{figure*}
%%%

In this section, we present the 16-site ED study, which served as a preliminary survey of the phase boundaries in a finite sized cluster.
The 16-site cluster is composed of two Bravais lattice points in the global $\hat{x}$ and $\hat{y}$ direction, and a single lattice point in the global $\hat{z}$.
{Figure} \ref{fig_ed_16_schematic} depicts both a schematic of the 16-site cluster and the precise locations of the sublattices with respect to the boundary conditions.

As described in \textit{Methods}, we perform the ED computation to obtain the following zero-temperature phase diagram in Fig. \ref{fig_ed_16_phases}, at zero magnetic field.
As seen, we confirm the existence of three distinct phases separated by phase boundaries.
The phase boundaries are characterized as a singular point in the 2$^{\rm nd}$ derivative of ground state energy with respect to the two coupling constants i.e. singularity in $\partial^2 E/\partial J_{\pm}^2$ and $\partial^2 E/\partial J_{\pm\pm}^2$. 
We compare the qualitative similarity of Fig. \ref{fig_ed_16_phases} to the classical phase diagram of Fig. \ref{fig_classical_phases} as well as the gMFT phase diagram in Ref. \onlinecite{sbl_gmft}.
Although the precise location of the phase boundaries is different when comparing ED to gMFT (or even classical) studies, this is merely a consequence of the finite-size effects of ED.

To understand the nature of these phases, we examine the static-spin structure factor for each of the phases.
We use the same definition of the static structure factor as in \textit{Methods},
\begin{align}
  S^{\alpha\beta}({\bm q}) =\sum_\mu S^{\alpha\beta}_\mu({\bm q})=  \frac{1}{N_s}\sum_\mu\sum_{i\in\alpha,j\in\beta}e^{-i{\bm k}\cdot({\bm R}_i-{\bm R}_j)}\langle{  \sfS}_i^\mu  {  \sfS}_j^\mu \rangle,
\end{align}
where $\mu$ sums over the three components $\{x,y,z\}$ of the pseudospin, $\alpha$ and $\beta$ are sublattice indices $\{0,1,2,3\}$, $N_s$ is the total number of sites, $i,j$ are site locations of sublattice $\alpha, \beta$, respectively; in the $N_s=16$ site cluster, there are four such $i,j$ locations each.
The wave number ${\bm k}$ is represented by using primitive reciprocal vectors ${\bm b}_i$ as
 ${\bm k}=\sum_{i=1}^3 q_i{\bm b}_i$. From this notation, we can easily find that the first Brillouin zone is for $-1/2<q_i<1/2$.
 For the 16 site cluster, there are two momenta points in the $k_x$ and $k_y$ direction namely that of $0$ and $\pi$, while the $k_z$ direction only has one momentum wavevector of $0$, since there is only one Bravais lattice point in the $z$ direction.

We present in Fig. \ref{fig_structure_factor_16} the static structure factor $S^{22}({\bm q})$ for each of the three regions of Fig. \ref{fig_ed_16_phases}, which provides information on the long-range correlation effects in the cluster.
In particular, the location of the peak structure provides information as to the nature of the multipolar order realized in the system.
For region I, we find intensity peaks at ${\bm q}=(0,0)$ and ${\bm q}=(\pi,\pi)$, which is a reflection of a lack of an ordering wavevector within the 16-site cluster.
This lack of order seems qualitatively consistent with a QSL phase.
On the other hand, phases II and III have a single peak located at ${\bm q}=(0,0)$, which validates a $\bm{q}=\bm{0}$ ordered state.

To distinguish phases II and III is, however, challenging as (described in \textit{Methods}) the expectation value of the local pseudospin moment in the absence of a symmetry-breaking magnetic field is always zero in ED.
By studying the pseudospin-pseudospin correlation function, we can fortunately demonstrate the consistency of these phases with the classical ordered 1D manifold and PC phases.
For instance, in phase II, the nearest-neighbour correlation is $ \langle \sfS_\alpha^{x(y)} \sfS_\beta^{x(y)}\rangle >0$ and $\langle \sfS_\alpha^z \sfS_\beta^z\rangle > 0^+$. 
Here $0^+$ indicates a positive, but order of magnetiude smaller number than $ \langle \sfS_\alpha^{x(y)} \sfS_\beta^{x(y)}\rangle$.
This indicates a ferro-like correlation in the $xy$ local moments (that is dominant over any $z$ component correlation), which is consistent with the 1D manifold of states.
Similarly, for phase III, the nearest-neighbour correlation is consistent with the PC phase.
Thus, we can reasonably claim that the 16-site ED phase diagram matches well with the expected phase diagram from gMFT. 
We use the 16-site ED phase diagram as a guide for the choice of parameters to use for the 32-site ED investigation of quantum spin ice.
In particular, in anticipation that the phase boundaries will likely shift, we choose $J_{\pm}$ and $J_{\pm \pm}$ to be deep in phase I (the likely quantum spin ice phase) and away from the phase boundaries.

%\vfill
%\pagebreak
\section{32-site ED Expectation Value of Quadrupolar Moments}
\label{app_32_ed_expectation}

\begin{figure}[h]
  \centering
  \includegraphics[width=\linewidth]{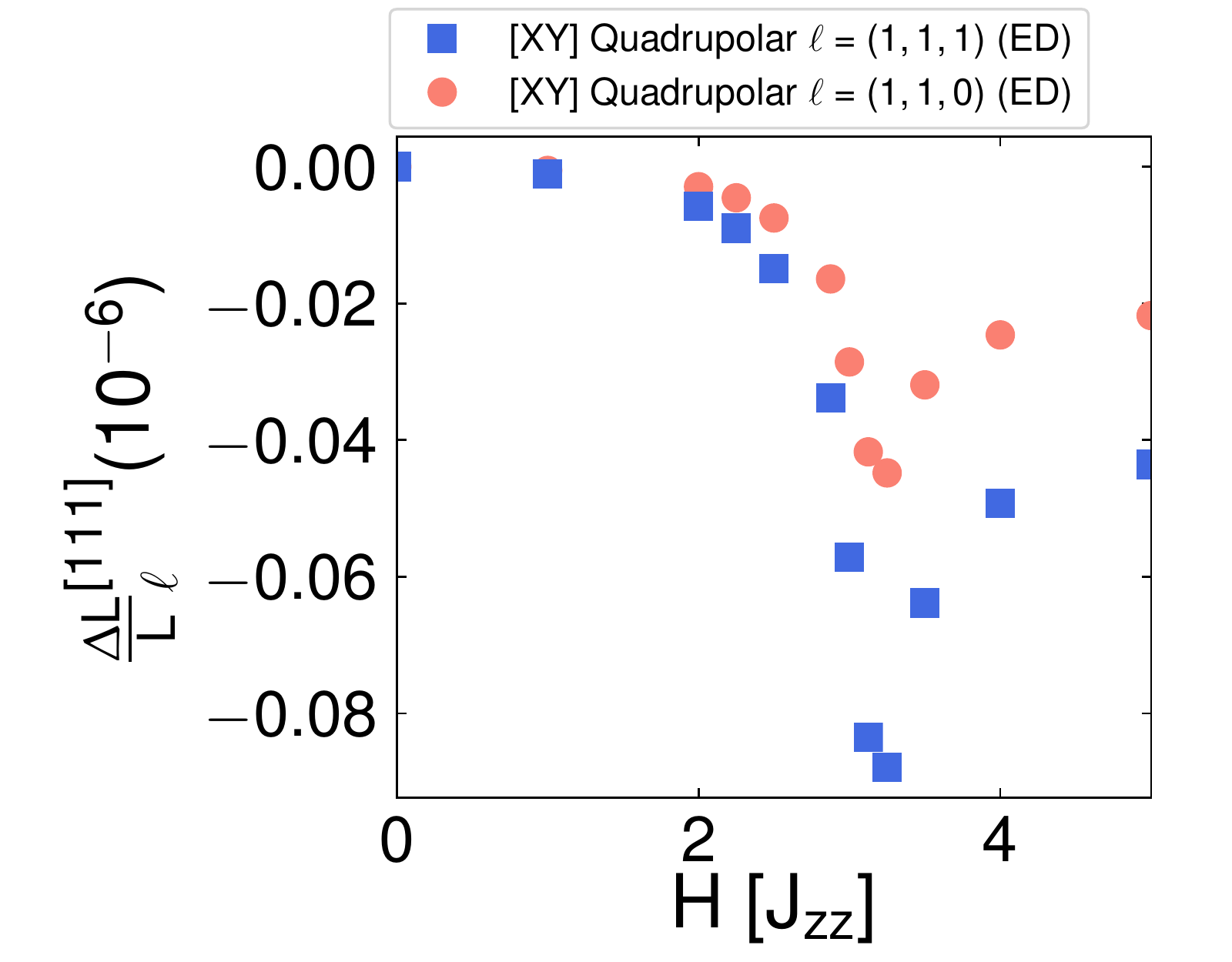}
  \caption{XY magnetostriction behaviour from directly obtained XY expectation value.}
  \label{fig_xy_expectation_direct}
\end{figure}

As described in \textit{Methods}, one can find the quadrupolar (XY) expectation value from (i) the `correlator' method method, or (ii) obtain the explicit expectation value directly.
In Fig. \ref{fig_xy_expectation_direct}, we use the directly obtained expectation value to plot the XY contribution to the magnetostriction under a [111] magnetic field, using the same coupling parameters detailed in Supplementary Information \ref{app_coupling_const_plots}.
%%%
%%%

\section{Irreps decomposition of NK and K magnetic orderings}
\label{app_irreps}

For completeness, we present the basis states of the various magnetic orderings of the Kramers and non-Kramers ions in Table \ref{TableI}.
Although the Kramers and non-Kramers ions have the same `types' of orderings, due to the difference in the symmetry transformations of $\overline{J_x J_z}$ and $J_x$ the SFM ($\rm{cAFQ_{L}}$) and PC ($\overline{\rm{cAFQ_{L}}}$) phases have different irrep labels.

\clearpage
\begin{turnpage}
\begin{table}[h!]
  \caption{Basis states for the magnetic orderings of Kramers and non-Kramers ions. Abbreviations: AIAO = all-in / all-out, PC = Palmer-Chalker, SI = Spin Ice, 1D manifold of states, SFM = Splayed Ferromagnet; MSI= multipolar spin ice, coplanar ferro-quadrupolar orders= $\rm{cAFQ_{L}}$, $\overline{\rm{cAFQ_{L}}}$, and $\rm{FQ_{L}}$ =  = ferro-quadrupolar. The two types of coplanar ferro-quadrupolar orders are related by a local $C_{4z}$ symmetry. The orderings are arranged into irreps of $T_d$ i.e. the pyrochlore point group.}
  \label{TableI}
  \centering
  \scalebox{0.8}{
  \begin{tabular}{c|c|c|c|c}\hline
    Name & Kramers case & Order & non-Kramers case & Order\\ \hline\hline
    $A_2$ & $M_{A_2} = \cfrac{1}{2}({\mathsf{S}}_{(0)}^z+{\mathsf{S}}_{(1)}^z+{\mathsf{S}}_{(2)}^z+{\mathsf{S}}_{(3)}^z)$ & AIAO
     & $M_{A_2} = \cfrac{1}{2}({\mathsf{S}}_{(0)}^z+{\mathsf{S}}_{(1)}^z+{\mathsf{S}}_{(2)}^z+{\mathsf{S}}_{(3)}^z)$ &AIAO
     \rule[-7mm]{0mm}{16mm}\\ \hline
     $E$ & $\overrightarrow{M}_{E} = \cfrac{1}{2}
     \begin{pmatrix}
      {\mathsf{S}}_{(0)}^x+{\mathsf{S}}_{(1)}^x+{\mathsf{S}}_{(2)}^x+{\mathsf{S}}_{(3)}^x \\
      {\mathsf{S}}_{(0)}^y+{\mathsf{S}}_{(1)}^y+{\mathsf{S}}_{(2)}^y+{\mathsf{S}}_{(3)}^y
    \end{pmatrix}
    $ & 1D Manifold & $\overrightarrow{M}_{E} = \cfrac{1}{2}
    \begin{pmatrix}
      {\mathsf{S}}_{(0)}^x+{\mathsf{S}}_{(1)}^x+{\mathsf{S}}_{(2)}^x+{\mathsf{S}}_{(3)}^x \\
      {\mathsf{S}}_{(0)}^y+{\mathsf{S}}_{(1)}^y+{\mathsf{S}}_{(2)}^y+{\mathsf{S}}_{(3)}^y
    \end{pmatrix} $ & $\rm{FQ_{L}}$
    \rule[-10mm]{0mm}{22mm}\\ \hline
    $T_2$ & $\overrightarrow{M}_{T_2} =
    \begin{pmatrix}
     \frac{1}{2}\left({\mathsf{S}}_{(0)}^y+{\mathsf{S}}_{(1)}^y-{\mathsf{S}}_{(2)}^y-{\mathsf{S}}_{(3)}^y\right) \\
     \frac{1}{4}(-\sqrt{3}{\mathsf{S}}_{(0)}^x-{\mathsf{S}}_{(0)}^y+\sqrt{3}{\mathsf{S}}_{(1)}^x+{\mathsf{S}}_{(1)}^y-
     \sqrt{3}{\mathsf{S}}_{(2)}^x-{\mathsf{S}}_{(2)}^y+\sqrt{3}{\mathsf{S}}_{(3)}^x+{\mathsf{S}}_{(3)}^y) \\
     \frac{1}{4}(\sqrt{3}{\mathsf{S}}_{(0)}^x-{\mathsf{S}}_{(0)}^y-\sqrt{3}{\mathsf{S}}_{(1)}^x+{\mathsf{S}}_{(1)}^y
     -\sqrt{3}{\mathsf{S}}_{(2)}^x+{\mathsf{S}}_{(2)}^y+\sqrt{3}{\mathsf{S}}_{(3)}^x-{\mathsf{S}}_{(3)}^y)
   \end{pmatrix}
   $ & PC & $\overrightarrow{M}_{T_2} =
  \begin{pmatrix}
   \frac{1}{2}\left({\mathsf{S}}_{(0)}^x+{\mathsf{S}}_{(1)}^x-{\mathsf{S}}_{(2)}^x-{\mathsf{S}}_{(3)}^x\right) \\
   \frac{1}{4}\left(-{\mathsf{S}}_{(0)}^x+\sqrt{3}{\mathsf{S}}_{(0)}^y+{\mathsf{S}}_{(1)}^x
   -\sqrt{3}{\mathsf{S}}_{(1)}^y-{\mathsf{S}}_{(2)}^x+\sqrt{3}{\mathsf{S}}_{(2)}^y+{\mathsf{S}}_{(3)}^x
   -\sqrt{3}{\mathsf{S}}_{(3)}^y\right) \\
   \frac{1}{4}\left(-{\mathsf{S}}_{(0)}^x-\sqrt{3}{\mathsf{S}}_{(0)}^y+{\mathsf{S}}_{(1)}^x
   +\sqrt{3}{\mathsf{S}}_{(1)}^y+{\mathsf{S}}_{(2)}^x+\sqrt{3}{\mathsf{S}}_{(2)}^y-{\mathsf{S}}_{(3)}^x
   -\sqrt{3}{\mathsf{S}}_{(3)}^y\right)
  \end{pmatrix}$ & $\rm{cAFQ_{L}}$
  \rule[-15mm]{0mm}{32mm}\\ \hline
  $T_{1,A}$ & $\overrightarrow{M}_{T_1,A} =\cfrac{1}{2}
  \begin{pmatrix}
   {\mathsf{S}}_{(0)}^z+{\mathsf{S}}_{(1)}^z-{\mathsf{S}}_{(2)}^z-{\mathsf{S}}_{(3)}^z \\
   {\mathsf{S}}_{(0)}^z-{\mathsf{S}}_{(1)}^z+{\mathsf{S}}_{(2)}^z-{\mathsf{S}}_{(3)}^z \\
   {\mathsf{S}}_{(0)}^z-{\mathsf{S}}_{(1)}^z-{\mathsf{S}}_{(2)}^z+{\mathsf{S}}_{(3)}^z
 \end{pmatrix}
 $ & SI & $\overrightarrow{M}_{T_1,A} =\cfrac{1}{2}
 \begin{pmatrix}
  {\mathsf{S}}_{(0)}^z+{\mathsf{S}}_{(1)}^z-{\mathsf{S}}_{(2)}^z-{\mathsf{S}}_{(3)}^z \\
  {\mathsf{S}}_{(0)}^z-{\mathsf{S}}_{(1)}^z+{\mathsf{S}}_{(2)}^z-{\mathsf{S}}_{(3)}^z \\
  {\mathsf{S}}_{(0)}^z-{\mathsf{S}}_{(1)}^z-{\mathsf{S}}_{(2)}^z+{\mathsf{S}}_{(3)}^z
\end{pmatrix}$ & MSI
\rule[-15mm]{0mm}{32mm} \\ \hline
$T_{1,B}$ & $\overrightarrow{M}_{T_1,B} =
\begin{pmatrix}
 \frac{1}{2}\left({\mathsf{S}}_{(0)}^x+{\mathsf{S}}_{(1)}^x-{\mathsf{S}}_{(2)}^x-{\mathsf{S}}_{(3)}^x\right) \\
 \frac{1}{4}\left(-{\mathsf{S}}_{(0)}^x+\sqrt{3}{\mathsf{S}}_{(0)}^y+{\mathsf{S}}_{(1)}^x
 -\sqrt{3}{\mathsf{S}}_{(1)}^y-{\mathsf{S}}_{(2)}^x+\sqrt{3}{\mathsf{S}}_{(2)}^y+{\mathsf{S}}_{(3)}^x
 -\sqrt{3}{\mathsf{S}}_{(3)}^y\right) \\
 \frac{1}{4}\left(-{\mathsf{S}}_{(0)}^x-\sqrt{3}{\mathsf{S}}_{(0)}^y+{\mathsf{S}}_{(1)}^x
 +\sqrt{3}{\mathsf{S}}_{(1)}^y+{\mathsf{S}}_{(2)}^x+\sqrt{3}{\mathsf{S}}_{(2)}^y-{\mathsf{S}}_{(3)}^x
 -\sqrt{3}{\mathsf{S}}_{(3)}^y\right)
\end{pmatrix}
$ & SFM & $\overrightarrow{M}_{T_1,B} =
\begin{pmatrix}
 \frac{1}{2}\left({\mathsf{S}}_{(0)}^y+{\mathsf{S}}_{(1)}^y-{\mathsf{S}}_{(2)}^y-{\mathsf{S}}_{(3)}^y\right) \\
 \frac{1}{4}(-\sqrt{3}{\mathsf{S}}_{(0)}^x-{\mathsf{S}}_{(0)}^y+\sqrt{3}{\mathsf{S}}_{(1)}^x+{\mathsf{S}}_{(1)}^y-
 \sqrt{3}{\mathsf{S}}_{(2)}^x-{\mathsf{S}}_{(2)}^y+\sqrt{3}{\mathsf{S}}_{(3)}^x+{\mathsf{S}}_{(3)}^y) \\
 \frac{1}{4}(\sqrt{3}{\mathsf{S}}_{(0)}^x-{\mathsf{S}}_{(0)}^y-\sqrt{3}{\mathsf{S}}_{(1)}^x+{\mathsf{S}}_{(1)}^y
 -\sqrt{3}{\mathsf{S}}_{(2)}^x+{\mathsf{S}}_{(2)}^y+\sqrt{3}{\mathsf{S}}_{(3)}^x-{\mathsf{S}}_{(3)}^y)
\end{pmatrix}
$ & $\overline{\rm{cAFQ_{L}}}$
\rule[-15mm]{0mm}{32mm}\\ \hline
  \end{tabular}
  }
\end{table}
\end{turnpage}
\clearpage
\global\pdfpageattr\expandafter{\the\pdfpageattr/Rotate 90}

\clearpage
\global\pdfpageattr\expandafter{\the\pdfpageattr/Rotate 0}

%\bibliography{qsi_magneto_bibtex}

\end{document}